\documentclass[manuscript,screen]{acmart}
\AtBeginDocument{%
  }

\setcopyright{rightsretained}
\copyrightyear{2022}

\usepackage{multirow}
\usepackage{multicol}
\usepackage{booktabs}
\usepackage{subcaption}
\usepackage{caption}
\usepackage{setspace}
\usepackage{amsmath}
\usepackage{tabularx}
\usepackage{balance}
\usepackage{textcomp}

\definecolor{midnightgreen}{rgb}{0.0, 0.29, 0.33}
\definecolor{darkpink}{rgb}{0.91, 0.33, 0.5}
\definecolor{darkmagenta}{RGB}{139, 0, 139}

\begin{document}


\title{ClueWeb22: 10 Billion Web Documents with Visual and Semantic Information}

\author{Arnold Overwijk}
\authornote{The first two authors contributed equally to this research.}
\email{arnold.overwijk@microsoft.com}
\affiliation{%
  \institution{Microsoft}
    \city{Redmond}
  \state{WA}
  \country{USA}
}
\author{Chenyan Xiong}
\authornotemark[1]
\email{chenyan.xiong@microsoft.com}
\affiliation{%
  \institution{Microsoft}
    \city{Redmond}
  \state{WA}
  \country{USA}
}
\author{Xiao Liu}
\email{xiliu2@microsoft.com}
\affiliation{%
  \institution{Microsoft}
  \city{Redmond}
  \state{WA}
  \country{USA}
}
\author{Cameron VandenBerg}
\email{cmw2@cs.cmu.edu}
\affiliation{%
  \institution{Carnegie Mellon University}
    \city{Pittsburgh}
  \state{PA}
  \country{USA}
}
\author{Jamie Callan}
\email{callan@cs.cmu.edu}
\affiliation{%
  \institution{Carnegie Mellon University}
  \city{Pittsburgh}
  \state{PA}
  \country{USA}
}

\begin{abstract}
ClueWeb22,  the newest iteration of the ClueWeb line of datasets, provides 10 billion web pages affiliated with rich information.
Its design was influenced by the need for a high quality, large scale web corpus to support a range of academic and industry research, for example, in information systems, retrieval-augmented AI systems, and model pretraining.
Compared with earlier ClueWeb corpora, the ClueWeb22 corpus is larger, more varied, of higher-quality, and aligned with the document distributions in commercial web search.  
Besides raw HTML, ClueWeb22 includes rich information about the web pages provided by industry-standard document understanding systems, including the visual representation of pages rendered by a web browser, parsed HTML structure information from a neural network parser, and pre-processed cleaned document text to lower the barrier to entry. 
Many of these signals have been widely used in industry but are available to the research community for the first time at this scale.
\end{abstract}
\maketitle

\section{Introduction}

Large scale web corpora are essential for the research and development of technologies such as information retrieval (IR), natural language processing (NLP), and deep learning.
Previous ClueWeb corpora, ClueWeb09~\citep{ClueWeb09} and ClueWeb12~\citep{ClueWeb12}, have been the standard web corpora for many research explorations during the last decade (e.g.,~\citep{clarke2009overview, clarke2012overview, yang2019xlnet}), but they are more than ten years old.
The web and users have changed, new research frontiers have emerged, and our web corpora must evolve with them.

Many important research topics in information retrieval are centered around web search.
As one of the most used AI applications, web search presents research challenges in user understanding, relevance modeling, indexing, serving efficiency, and many more. All of them rely on a large scale, realistic web corpus to reflect the data distribution and the realistic challenges of web search.
Perhaps due to the lack of updated web corpora, many recent IR benchmarks use the derived MS MARCO corpus~\citep{bajaj2016ms} which was sampled from the top retrieved passages of a Bing question answering system nearly six year ago, which is very different from the distribution of the web.

An emerging usage of web corpora is in retrieval-augmented AI systems,
for example, to provide documents that serve as evidence for question answering~\citep{chen2017reading}, grounding information to improve language generation~\citep{lewis2020retrieval},
and enriched contexts for language models~\citep{guu2020retrieval}.
A large fraction of these systems use Wikipedia as the retrieval corpus, which provides high quality but only a fraction of human knowledge.
In comparison, the web is a much richer source of human knowledge, but obtaining a high quality large scale web corpus is costly and challenging in most environments~\citep{piktus2021web}.

Another increasingly important usage of web corpora is to provide data for neural model pretraining.
As the amount of network parameters has grown from hundreds of millions to trillions in the past four years~\citep{devlin2019bert, fedus2021switch}, their pretraining also requires significantly more data~\citep{hoffmann2022training}.
Many sought more pretraining data from the web. 
One approach is to sift CommonCrawl snapshots to find relatively higher quality web pages, for example, as the C4 dataset was cooked to pretrain T5~\citep{raffel2019t5}. 
Web corpora cooked from CommonCrawl provide sufficient quantity but not necessarily the highest quality. 
A new series of efforts have curated better web corpora to empower language model pretraining using proprietary resources, such as MassiveText~\citep{rae2021scaling} in DeepMind and the pretraining corpus of GaLM~\citep{du2021glam} and PaLM~\citep{chowdhery2022palm} in Google.
These higher quality web corpora contributed significantly to the effectiveness of pretrained models~\citep{du2021glam}, but were only available in a few places, perhaps exclusively to companies with a commercial web search business.

\begin{table}[t]
    \centering
    \small
        \caption{Size and sampling distribution of ClueWeb22. We follow the previous ClueWeb corpora to designate official subsets of web pages: Category B $\subset$ Category A 
  $\subset$ Category L.}
    \label{tab:size}
    \begin{tabular}{lrrl}
    \toprule
    \textbf{Category} & \textbf{\#Pages} & \textbf{\#Tokens}  & \textbf{Sampling Distribution} \\ \hline
        ClueWeb22-B & 200M & 696B 
        & From Most Popular Web Pages (``Super Head'')
        \\ \hline
        ClueWeb22-A & 2B & 6.1T
        & From Pages also Frequently Visited by Users (``Head'') \\ \hline
         ClueWeb22-L & 10B & 16.7T 
         & Mixed Head-Tail Pages (``Head and Tail'')\\ 
         \bottomrule
    \end{tabular}
\end{table}

These research needs are among many we would like to support with ClueWeb22, the large scale, industry-quality, web corpus that is now available to the research community.
The construction of ClueWeb22 emphasizes the following goals: 1) \textit{Real Distribution:} to reflect the distribution of the web in real-world scenarios; 
2) \textit{Large Scale Quality Content:} to provide clean web content at large scale with high quality; 3) \textit{Rich Information:} to share information beyond raw text on web pages that are widely used in industry, but previously unavailable for academia.

\textbf{Real Distribution.} 
The web pages from ClueWeb22 come from the web discovered by the crawler of a commercial search engine, which provides a comprehensive representation of the web.
Then its web pages are sampled from all these web pages discovered and indexed by the search engine, following the distribution of web search. 

Specifically, a page more likely to satisfy potential information needs from search engine users has a higher probability to be included in ClueWeb22. 
In total, we sampled 10 billion web pages from the indexed corpus and grouped them into three ``categories''. As listed in Table~\ref{tab:size}, these categories follow the tradition of ClueWeb09 while also closely mimicking real scenarios used in web search. 
\begin{enumerate}
    \item ClueWeb22-B (Category B) approximates the super head scenario, the most frequently visited part of the web, e.g., the main parts of Wikipedia, news websites, and other top internet domains.  
    \item ClueWeb22-A (Category A) approximates the general part of the web regularly visited through search. It includes two billion web pages and covers most notable URL domains. 
    \item ClueWeb22-L (Category L) introduces the tail part of the web into the collection, with a total of ten billion web pages sampled from the index after spam and adult content filtering.  
\end{enumerate}
The three categories provide different trade-offs of quality and coverage. Each one is a subset of the larger ones, B $\subset$ A $\subset$ L, providing three choices for different research focuses.  
The size and the information included in categories of ClueWeb22 are decided to balance the information richness and the distribution cost.
A too large dataset may result in restricted access due to high distribution cost.

The details of the sampling process is described in Section~\ref{sec:construction}.
In Section~\ref{sec:stats} we analyze the properties of web pages in ClueWeb22, which closely resemble the distribution of the web visited from the web search engine, e.g., about 40\% web pages in ClueWeb22 are in English and the rest are in other languages.

\begin{table*}[t]
    \centering
        \small
            \caption{Information of web pages included in ClueWeb22. This information was obtained by state-of-the-art production-quality pipelines. We balance the availability of information in each ClueWeb22 category based on the eventual data size, which determines the distribution and storage cost.}
    \label{tab:feature}
    \resizebox{\textwidth}{!}{
    \begin{tabular}{lll}
    \toprule
    \textbf{Information} & \textbf{Categories} &  \textbf{Description} \\ \hline
    Raw HTML & A, B & The original, unprocessed HTML of the web page\\
    Clean Text & A, B, L & Primary text content, i.e., without headers, footers, side bar, navigation panel, etc. \\ 
     
    Semantic Annotations & A, B, L & Annotations of content structure: title, section headings, paragraphs, lists, and tables \\
    Anchor Text \& Inlink & A, B, L & The inlinks of a web page and their affiliated texts extracted from ClueWeb22-A \\
    Anchor Text \& Outlink & A, B & The outlinks of a web page and their affiliated texts extracted from its HTML \\
    VDOM Features & A, B & The visual representation features of the content, e.g., location and size of each text piece \\
    Rendered Visual Page & B & The screen shot of the rendered web page in a web browser \\
     Language Tag & A, B, L & The language of the main content, detected by an updated version of BlingFire~\citep{BlingFire} \\
    Topic Tag & A, B, L & The category classified by a supervised neural classifier \\ 
         \bottomrule
    \end{tabular}
    }

\end{table*}

\textbf{Large Scale Quality Content.}
To extract high quality content from raw HTML pages requires access to information and engineering efforts that are rarely available outside large commercial systems.
Besides raw HTML data, ClueWeb22 also includes parsed clean text from the search engine's production-quality content understanding system, which includes engineering and research advancements accumulated through years of dedicated work.
The full content extraction pipeline is described in Section~\ref{sec:parsing}, which involves a web browser to render web pages and a deep neural network to classify HTML elements.

To lower the entry barrier, we provide the extraction results of several notable fields of the web page, for example, title, primary text content, tables, lists, and anchor links.
In addition, all the information used in the parsing API, and the API toolkit itself, are made available with ClueWeb22. Researchers can customize the content extraction to obtain information specific to their needs, for example, image-text pairs for vision-language pretraining.

\textbf{Rich Information.}  We include in ClueWeb22 a variety of information produced by our document understanding system, listed in Table~\ref{tab:feature}. 
Much of this information provides important signals for industry systems, but was not available to the research community. 
Making this information available to the research community via ClueWeb22 could facilitate the technology development towards directions important to real-world systems, but previously overlooked in academia.
For example, the semi-structured content, such as tables and lists, are critical parts for modern QA systems, but most previous research is restricted to Wikipedia due to limited data availability~\cite{ma2021open}; the visual presentation of documents is widely used in practical document understanding systems but less studied in academia. There were no such data available at this scale~\cite{openkpe}.

Starting August 2022, ClueWeb22 is available for research usage, following the licensing and distribution processes used with previous ClueWeb datasets.
The scale, distribution, quality, and rich information of ClueWeb22 make it the only web corpus of its kind widely available for research. 
To the best of our knowledge, this is as close as it gets to the real web corpora used in cutting-edge industry systems.
Previously this type of data was only available in the private sector, making it a privilege for a few places with certain data accesses.
We believe the release of ClueWeb22 levels the playing field for the research community, enables more research explorations, and will facilitate future technology advancement in many areas.

In the rest of this paper, Section~\ref{sec:construction} describes the construction of ClueWeb22;
Section~\ref{sec:parsing} discusses the content extraction pipeline, followed by analysis of corpus properties in Section~\ref{sec:stats}.
Then we provide some comparison with CommonCrawl in Section~\ref{sec:cc} and discuss related datasets in Section~\ref{sec:related}.
After that, we briefly share the licensing and distribution in Section~\ref{sec:licensing} and conclude in Section~\ref{sec:conclusion}.

\section{Corpus Construction}
\label{sec:construction}

The first design principle of ClueWeb22 is to reflect the distribution of the web, which itself can be defined in different ways. For example, the importance of a web page can be derived from how many other pages link to it~\citep{page1999pagerank}, how it is discussed on Reddit~\citep{Gokaslan2019OpenWeb}, or how frequently web users browse it. In ClueWeb22, we formulate the distribution of web pages from the web search perspective, by modeling the probability of each web page based on how likely it would satisfy web search users information needs, i.e., receiving a satisfied user click.

\textbf{Web Page Sampling.} We leverage a production model  to predict the likelihood of a web page receiving a satisfied click from any search users via any queries.
The prediction model leverages a wide range of information available in web search, for example, web graph connectivity, URL domain, document content, and page structure, to name a few. The model is trained  by user clicks in web search and assigns a query-independent click likelihood to each web page in the search index.
We use this predicted click likelihood as the importance score of the web page. With it, the web pages discovered by the crawler can be roughly grouped into four groups:
\begin{itemize}
    \item \texttt{Super Head}: The most popular web pages visited through web search, such as the content pages of Wikipedia, popular news websites, and top domains people visit in their daily life;
    \item \texttt{Head}: The regularly visited part of the web, where most search traffic lands;
    \item \texttt{Tail}: The diverse part of the web still regularly visited by users with specific needs;
    \item \texttt{Super Tail}: The majority of the web discovered by crawlers but barely visited by users.
\end{itemize}

We sampled ClueWeb22 from the first three groups via the following steps.
\begin{enumerate}
    \item Uniformly sample 200 million web pages for Category-B (ClueWeb22-B) from the Super Head with emphasis on covering web pages with highest predicted click likelihood;
    \item Uniformly sample 1.8 billion web pages from the Head and combine them with the first 200 million to form Category-A;
    \item Uniformly sample the rest 8 billion web pages from the tail group and mix them with ClueWeb22-A to form Category-L .
\end{enumerate}

Before sampling, we manually excluded some websites that we considered to be mainly about personal information. Standard spam/adult filters were applied before sampling.
We decided not to include the Super Tail part of the web as our manual examination found the quality of web pages there quite low. 
To include sufficient useful information from the Super Tail, the scale of the corpus would exceed a reasonable distribution cost.
ClueWeb22-L already includes tail web pages less explored in previous web corpora.  We consider it a good balance of quality and coverage.

\begin{table}[t]
    \centering
    \small    
     \caption{English Wikipedia pages included in three categories of ClueWeb22, manually picked from one thousand random samples.}
    \label{tab:wikiurl}
    \resizebox{\linewidth}{!}{
    \centering
    \begin{tabular}{l|l}
    \hline
     \textbf{ClueWeb22-B} & \textbf{Page Type}\\
     \url{https://en.wikipedia.org/wiki/Super_Bowl_XXXVIII}  & Entity\\
     \url{https://en.wikipedia.org/wiki/Chevrolet_Corvette_(C8)}  & Entity \\
     \url{https://en.wikipedia.org/wiki/Discourse}  & Entity \\ \hline
\textbf{ClueWeb22-A}  & \textbf{Page Type}\\
\url{https://en.wikipedia.org/wiki/1897_in_film} & List \\
\url{https://en.wikipedia.org/wiki/2019_FFA_Cup_Final} &  Entity \\
\url{https://en.wikipedia.org/wiki/Category:Public_administration_schools_in_the_United_States}  & Category \\
\hline
\textbf{ClueWeb22-L}  & \textbf{Page Type} \\
\url{https://en.wikipedia.org/wiki/Template\%3ANeighbourhoods_in_Kolkata} & Edit Template\\
\url{https://en.wikipedia.org/wiki/Category:Sports_leagues_established_in_1990} & Category \\
\url{https://en.wikipedia.org/wiki/Talk:The_Shoes_of_the_Fisherman} & Wiki Project \\
         \hline
    \end{tabular}
   }
\end{table}

\textbf{Examples of Web Pages with Different Sampling Importance.} 
The query-independent click prediction system has quite high prediction accuracy. It has been used for awhile in many real world applications. To show how the documents in these categories align with our intuitions of  web page popularity, we use Wikipedia as an example and list some random Wiki URLs from ClueWeb22-B, ClueWeb22-A, and ClueWeb22-L in Table~\ref{tab:wikiurl}.

The total number of discovered URLs from Wikipedia.org is quite large, way bigger than the number of pages  in official Wikipedia dumps. Among these Wiki URLs, the entity pages, the typical content pages we read on Wikipedia, only form a tiny fraction. A notable amount of Wiki URLs are redirects, categories, and functional pages for Wikipedia construction efforts. Many are not informative.
As shown in Table~\ref{tab:wikiurl}, the importance prediction model naturally separates the Wiki URLs into different ClueWeb22 categories. 
Those in ClueWeb22-B are mostly entity pages, the part search users visit the most.
Category and List pages, although less informative but still including meaningful information, start to appear in ClueWeb22-A.
The functional pages that are for editors, not web users, are mostly assigned to ClueWeb22-L.

\section{Web Page Understanding and Content Extraction}
\label{sec:parsing}

An often necessary step working with web pages is to extract the content from raw HTML, such as main body text, tables, lists, hyper links, and multi-media content. Content extraction is challenging, because the web is sophisticated. There are lots of diversity from both the user perspective (layout of the page) and the system perspective (the underlying codes to present the web page).

If restricted to only using the HTML of web pages, it is even more difficult to extract clean content from them. The HTML alone only provides a partial view of the web page. The page layouts, and sometimes even content, require downloading and dynamically rendering with additional resources (e.g. CSS and JavaScript). For example, without actually rendering the page, it is hard to tell the visibility of text pieces, their location in the page, and the organization of structural contents.

On average, there are 50+ secondary URLs needed to render one web page.
Not only downloading them imposes significant cost, rendering the web page and executing the secondary contents is also non-trivial. It is effectively running an actual web browser.
As a result, many open-source web page parsing tools restrict their operations to static HTML, limiting their ability to extract the exact content a web page displays to users.

\begin{figure}
    \centering
    \includegraphics[width=1\textwidth]{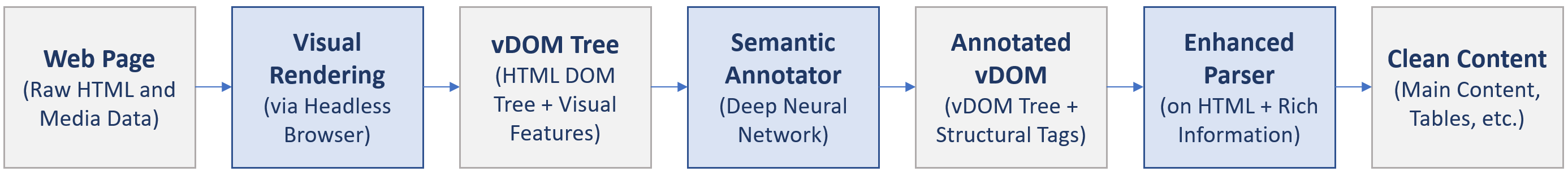}
    \caption{The system pipeline to extract content from ClueWeb22 web pages. The extracted Clean Content is available for all ClueWeb22 pages which is a good starting point for common applications.  The Annotated vDOM is available in ClueWeb22-A to support customized content extraction for more specific needs.}
    \label{fig:plm-frame}
\end{figure}

In this section we describe the industrial-strength web page understanding system used to extract content from ClueWeb22. As illustrated in Figure~\ref{fig:plm-frame},
it enhances traditional HTML parsing with the visual representations of the page elements and a Transformer-based model.
The full system includes three main components, starting from  a \textit{visual render} (Sec.~\ref{subsec:virtual_render}), that records the visual appearance of HTML tree nodes, then a \textit{semantic annotator} (Sec.~\ref{subsec:annotation}), that predicts the content categories of HTML elements, and finally an \textit{enhanced parser} (Sec.~\ref{subsec:parsing}), which extracts content from HTML aided by visual information and model predictions.

\subsection{Visual Render}
\label{subsec:virtual_render}

There are many different ways for web designers to present content to their visitors. For example, the behavior of any HTML tag (e.g. $<$h1$>$) can be flexibly customized through CSS and JavaScript. 
While there are standard guidelines in web page construction, in practice many different ways are used to accomplish the same presentation. 
Given the high variant nature of web programming, the ultimate way to determine which part of the page includes useful content is based on their design target: the presentation of web pages in web browsers.

The web page understanding system used upon ClueWeb22 is built towards this fundamental solution.
When a web page is crawled, it fetches both the raw HTML and its secondary resources. It then uses the full information to render the web page in a headless browser. 
The browser uses a standardized desktop environment of 1024 pixels width and records the most useful visual features for each HTML DOM node.
In total it records 30 visual features for each HTML element node\footnote{In ClueWeb22, element nodes refer to those with HTML tags, e.g., $\langle$html$\rangle$, $\langle$body$\rangle$, $\langle$div$\rangle$, and $\langle$p$\rangle$.}.  The full list of visual features is in Table~\ref{tab:visual}. Text nodes---the text pieces inside element nodes---inherit visual features from their parent element nodes.

As shown in previous research~\citep{openkpe, ainslie2020etc}, an earlier version of these virtual features, though released at a much smaller scale, contributed significantly to the accuracy of some document understanding tasks, for example, in web page keyphrase extraction~\citep{ainslie2020etc}. 
To balance the total size of ClueWeb22 and adhere to the distribution cost constraints, we include HTML DOM tree enhanced with visual features for ClueWeb22-A web pages; the screen shots (1024 width and max height of 6144) of web pages rendered by the browser are available in ClueWeb22-B.

\begin{table}[t]
    \centering
    \small
     \caption{Visual Features. All features are extracted at the token level and the parent block level (the parent node of the text in the HTML DOM tree).  We merge the description of features in the same group, separated by ``/''. Features about position and size are measured at pixel level in the rendered web page.\label{tab:visual}} 
    \begin{tabular}{l|l} \hline 
        \textbf{Name} & \textbf{Description} \\ \hline
        X/Y position &  Horizontal/vertical position  \\
        Width/Height & Width/height in rendered pixels 
        \\ 
        Offset left/top & Left/top offset relative to the parent element \\ 
        Offset width/height & Element's width/height including padding and borders \\ 
        Client left/top & Element's left/top border width\\ 
        Client width/height & Element's width/height including padding \\ 
        Font color: Transparent/Red/Green/Blue & The value of font color in the four dimensions\\ 
        Font weight/size & Font weight (or boldness)/size \\ 
        Font italic style & Whether Font in italic style\\ 
        Text decoration style & Decoration added to text such as underline, overline \\ 
        List style type & List-item marker type for a list \\ 
        Display & Display behavior of an element, such as none, inline, block \\ 
        Cursor & Mouse cursor display type when pointing over an element \\ 
        Line Height & Line box height \\ 
        Text transform & Text capitalization style \\ 
        Opacity & Opacity level\\
        Border style Left/Top/Right/Bottom &  The border style of each side\\ 
        \hline 
    \end{tabular}
\end{table}
\subsection{Semantic Annotator}
\label{subsec:annotation}

With the visual enhanced DOM tree (vDOM), the next step in the pipeline is to use a neural
\textit{semantic annotator} to classify vDOM tree nodes into target content groups. In next part of this section we discuss the annotation task, the model architecture, and the training labels of the semantic annotator.

\textbf{Semantic Annotation Task} is to predict whether a node in the HTML DOM tree belongs to a set of predefined categories. In ClueWeb22 the model uses the following six categories:
\begin{enumerate}
    \item \texttt{Title}: The title of the document content, which may be different from the HTML $<$title$>$ tag.
    \item \texttt{Primary Content}: The main content of the web page. Formally a piece of text is primary content if it is the main information for visitors to consume. This excludes elements that occur on other pages of the same site, such as headers, footers, and navigation, as well as elements that change with page reloading, e.g., advertisements. In addition, we remove elements that are not the core content, for example, the comment section of a blog site, where the blog is the core content.
    \item \texttt{Heading}: The heading of each section in the primary content. 
    \item \texttt{Paragraph}: The natural language paragraphs of the primary content.
    \item \texttt{Table}: Content tables in the primary content, grouped in tabular format. This is determined by the page presentation to users, regardless their organization in the static HTML or dynamic scripts.
    \item \texttt{List}: Content lists in the primary content, grouped in the list format.
\end{enumerate}

Because ClueWeb22-L does not include the HTML or vDOM due to size constraints, we include additional format annotations in the clean content to convey the structure information. These format annotations are derived from annotator predictions.
\begin{enumerate}
\setcounter{enumi}{6}
    \item \texttt{Table Row}: A row that contains multiple cells in a table.
    \item \texttt{Table Cell}: The element storing one content unit in a table.
    \item \texttt{Table Header}: The row or column header that contains labels for content in a table. 
    \item \texttt{Table Caption}: The description or summary of table content. 
    \item \texttt{List Item}: The element storing one content unit in a list.
    \item \texttt{HTML Title}: The page title displayed in the browser tab. 
    \item \texttt{Invisible Text}: Text that is invisible when rendered in the browser, i.e. those with zero opacity or less than two pixel in both width and height.
\end{enumerate}

The prediction of Title and Primary Content is applied on non-empty text nodes in the vDOM.
Heading and Paragraph predictions are on element nodes. The classification of Table and List are performed on specific element nodes, $\langle table\rangle$ for table,  $<$ol$>$, $<$ul$>$, and $<$dl$>$ for list. 
The prediction is carried as a multi-label task. A node can belong to multiple categories, for example, both being a Table and Primary Content, or none of them.

\textbf{Model Architecture.} 
The classification is conducted by a hierarchical Transformer network applied on the DOM tree. 
At its first level, the network leverages a pretrained multi-lingual BERT-style model to produce a text representation for each target DOM tree node. Then it concatenates the text representation of a node with the  projected representation of its visual features to form visual-enriched embeddings.
At the second level, another shallow Transformer network takes the input sequence of  visual-enriched embeddings from all nodes, produces contextualized representations, and performs multi-label classification on each node. The full hierarchical Transformer is trained end-to-end with the first level network initialized by XLM-R~\citep{conneau2019unsupervised} and the second level from scratch. 

\textbf{Weak Supervision Labels.} Manually labeling the role of vDOM nodes is quite difficult. We instead train with a large amount of weak supervision labels, accumulated through years of engineering and system iterations. The weak supervision signals come from the predictions of rule-based content extractors and classic feature-based models. 

The trained semantic annotator is applied to the entire ClueWeb22 corpus and the predicted labels for all ClueWeb22 documents are included in ClueWeb22.

\subsection{Enriched Parser}
\label{subsec:parsing}

The visual features and semantic annotations are linked to the HTML DOM tree nodes using a \textit{node id} attribute as the unique identifier. This can be used to enhance standard HTML parsing systems with visual and semantic annotation information.
In the rest of this section, we describe the parser used to generate the ClueWeb22 clean text corpora. It is a vanilla way to extract content. We consider it sufficient for many applications and suggest use it as the starting point to build more dedicated parsers, if needed.

In general, the parser includes into two steps: the first is to construct the vDOM tree from raw HTML; the second is to extract target content fields from the tree.

\textbf{vDOM Tree Construction} is to parse the HTML sequence into a tree.  One can leverage a typical HTML parser and use the injected ``node id'' attribute to align the parsed tree nodes with their visual features and semantic annotations. We used Beautiful Soup\footnote{\url{https://pypi.org/project/beautifulsoup4/4.9.3/}} for ClueWeb22.

\textbf{Target Field Extraction} is to extract content for the annotated document fields by traversing the parsed tree. Some of document fields are straightforward. On them We followed the common practices. Some of them require enhanced treatments and we discuss these in more detail.

\texttt{Primary Content.} All the texts under primary content tags are extracted to form the primary content. A naive concatenation of the text sequences from multiple tree nodes, however, does not yield well-formatted text passages. There is often a noticeable amount of extra spaces, tabs, and line breakers, which destroy the text flow. Also, quite frequently web pages include text pieces invisible to the readers. The invisible texts are a mix of meaningful content, functional text, and spam.
Identifying them from raw HTML DOM tree without visual features is nearly impossible.

To produce well-formatted primary content, we format the extracted text pieces by considering DOM tree hierarchy relationships, texts and HTML tags, visual features, and primary content annotations. The exact logic can be found in our open source repository. 
To handle the invisible texts, we use visual features (opacity, width, and height) to decide text visibility and include ``invisible'' annotation tags in the clean contents. One can decide whether to  filter them when using ClueWeb22 based on specific needs.

\texttt{Table and List.} Tables and lists include multiple elements, such as cells and items. Their semantic annotations are predicted at the element node level.
Our extraction system iterates through the vDOM tree and extracts the children elements of tagged element nodes: table header, table caption, table row, table cell, and list item elements. It then combines them into a table or list.

Besides extracted content, we also release the intermediate information, including the rendered pages (as JPEG images), virtual features, and semantic annotations to support more research explorations. 
For example, one can extend our open sourced extraction system to extract other fields such as image and hyperlinks, or develop one's own content extraction logic with the benefit of visual features and semantic annotations.
It is also possible to leverage the annotations in ClueWeb22 as weak supervision signals to train new annotation models.

\subsection{Anchor Graph Construction.} 

The hyperlinks between web pages and the affiliated texts, often referred as the anchor graph, provide valuable information for many applications. 
The link structure is a significant source of signal for web page importance estimation, for example, in PageRank~\citep{page1999pagerank} and in the click likelihood prediction model used to sample ClueWeb22. 
The hyperlink texts affiliated with inlinks, the ``anchor texts'', are important contexts for the corresponding documents in search~\citep{metzler2009building}. They are commonly viewed as close approximations of search queries and used as weak supervisions for search models~\citep{zhang2020selective}. 

The hyperlink information is available in the HTML data of ClueWeb22-A. One can leverage standard parsers to extract the hyperlinks and construct the anchor graph.
To facilitate research explorations and standardize evaluations, 
we pre-construct the anchor graph using ClueWeb22-A HTML vDOM and include it in the ClueWeb22 distribution.

Following the standard approach, we first extract the hyperlinks and their associated texts for each document in ClueWeb22-A (outlinks) and then merge them to get inlinks for each document. By design, outlinks can point to a non ClueWeb document but inlinks all come from ClueWeb22-A pages. 
If a page pointing to an outlink page multiple times, we randomly keep one of the hyperlinks. At most one thousand inlinks and outlinks are kept per page. 
The down sampling is done first on outlinks and then on inlinks, making the latter a subset of the former.

In addition, we record whether a hyperlink comes from the header/footer of its origin web page. Hyperlinks in the header/footer of a web page are often functional, e.g. ``homepage'', ``contact us'', which may not be informative.
We keep the hyperlinks that come from the same domain but one can filter them out, for example, to avoid self-voting when calculating PageRank.

\begin{figure}[t]
 \centering
          \begin{subfigure}[t]{0.3\textwidth}
         \centering
         \includegraphics[width=\textwidth]{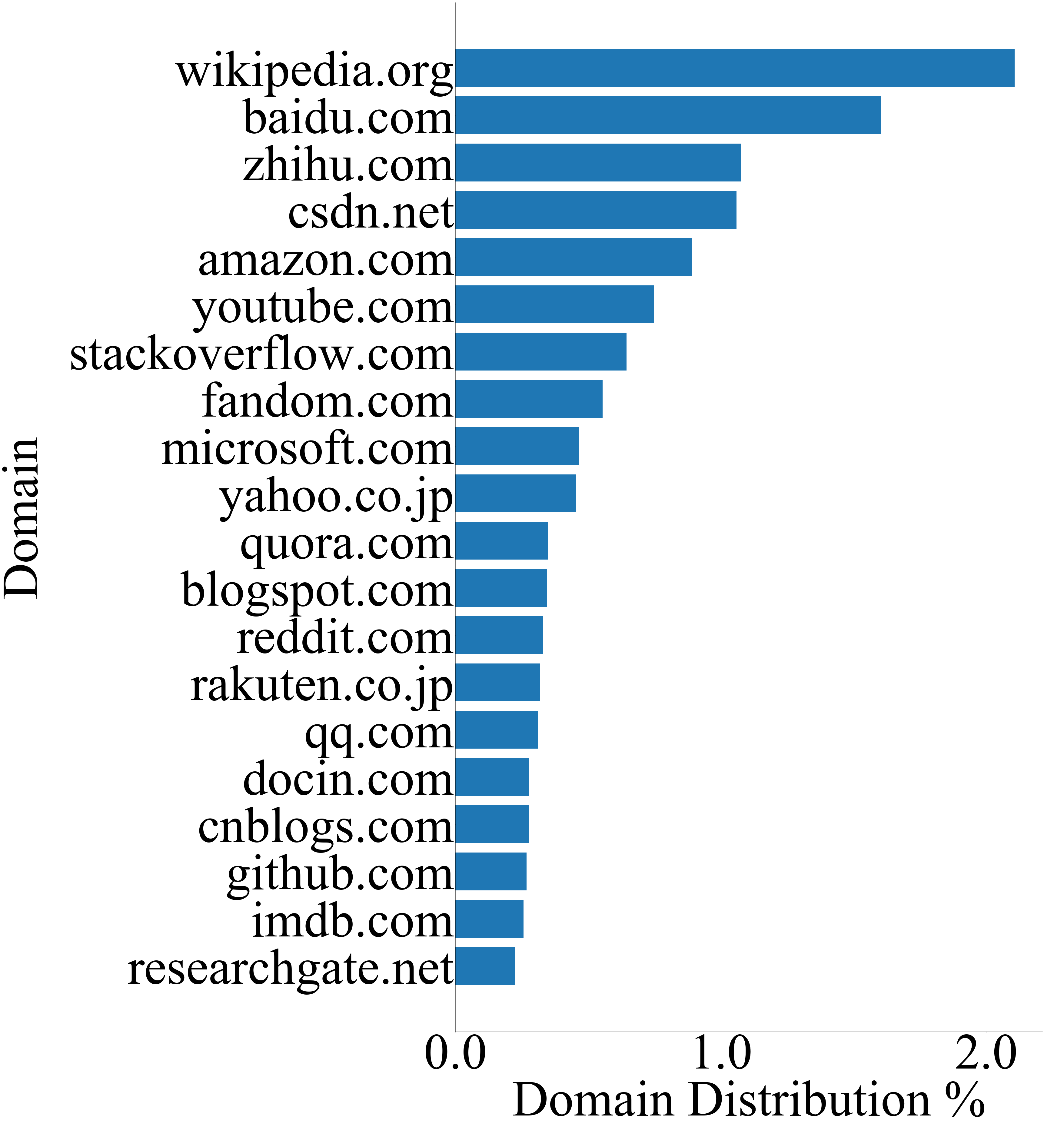}
         \caption{ClueWeb22-B.}
     \end{subfigure}
     \begin{subfigure}[t]{0.3\textwidth}
         \centering
         \includegraphics[width=\textwidth]{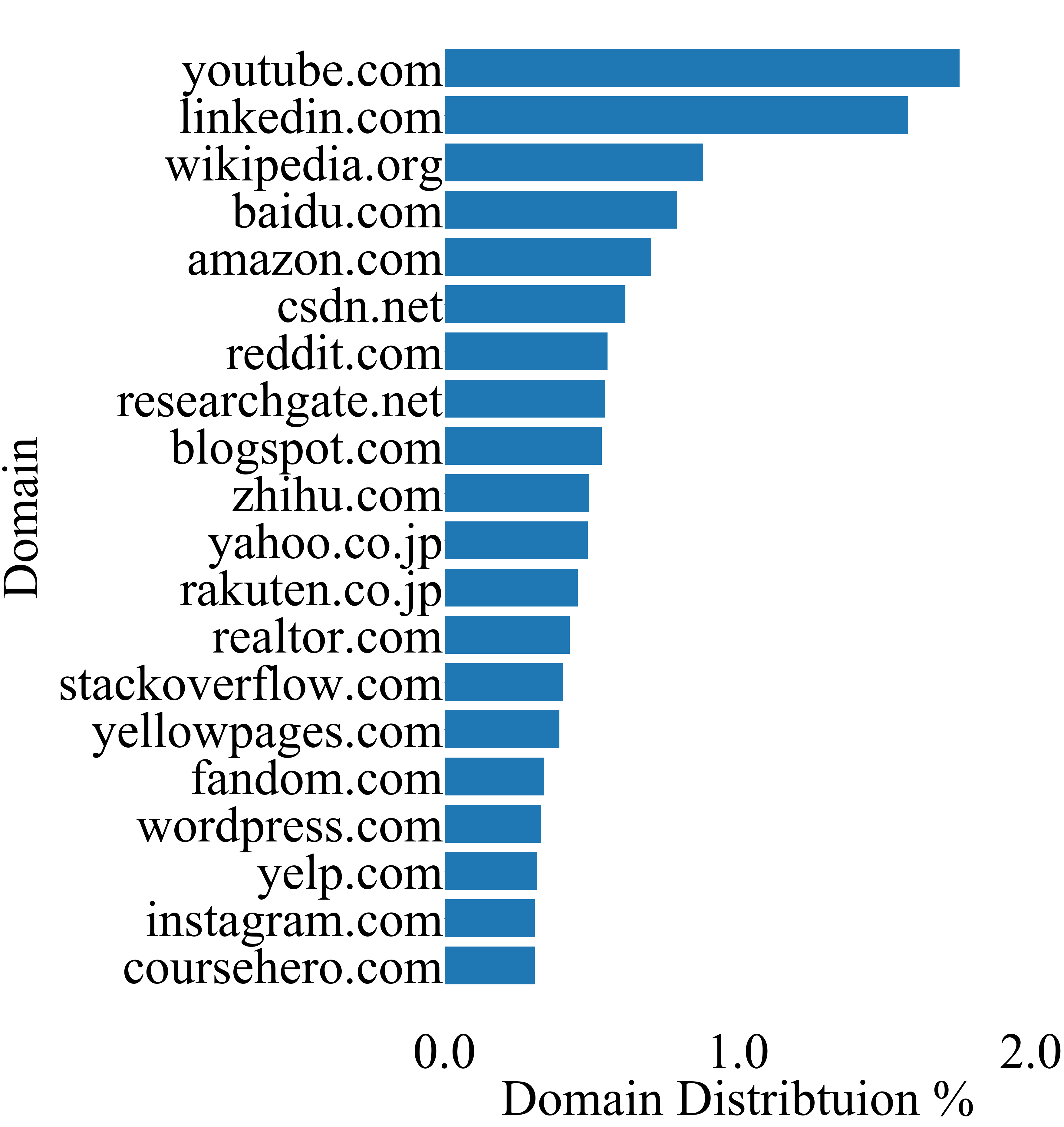}
         \caption{ClueWeb22-A.}
     \end{subfigure}
          \begin{subfigure}[t]{0.3\textwidth}
         \centering
         \includegraphics[width=\textwidth]{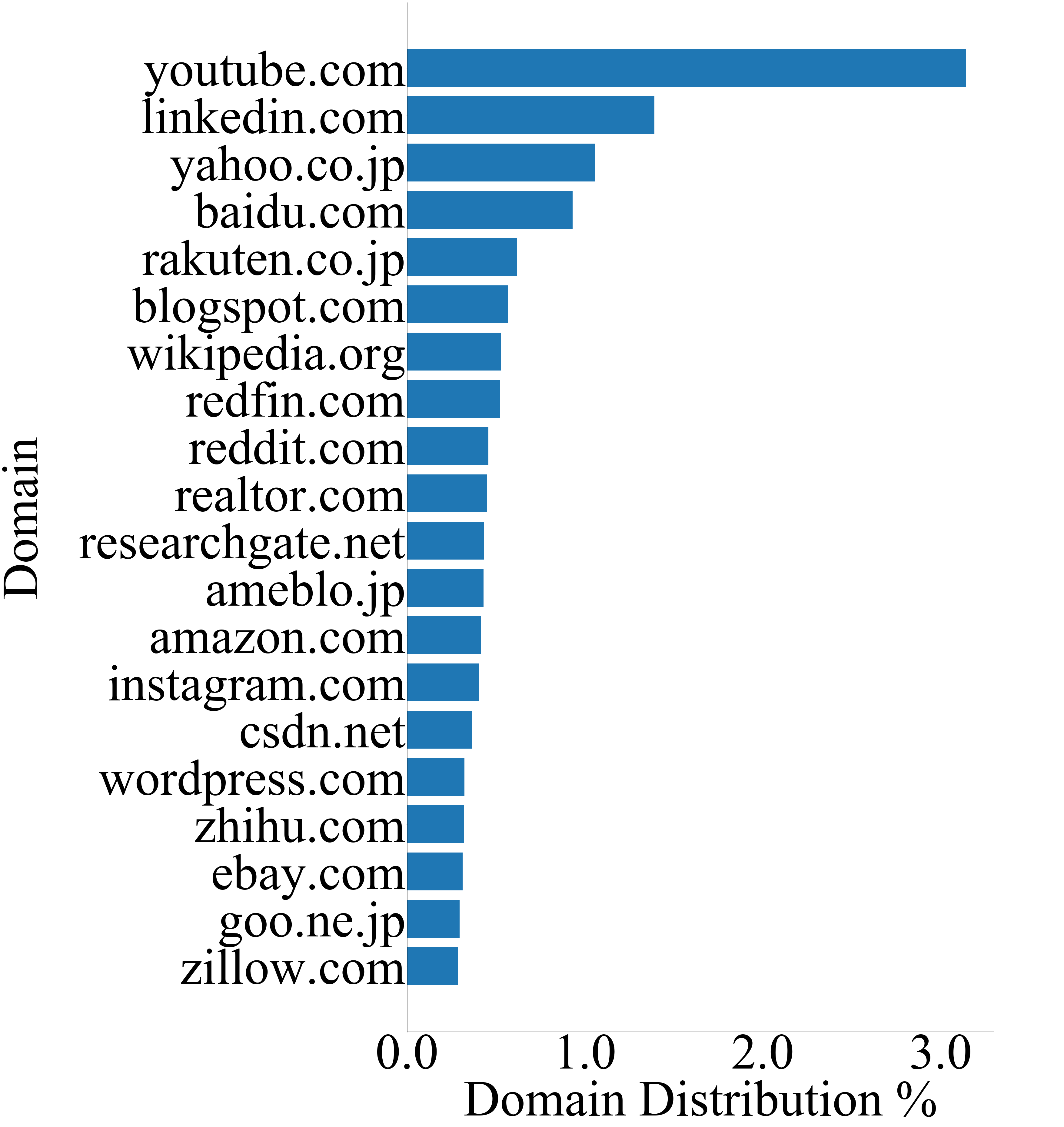}
         \caption{ClueWeb22-L.}
     \end{subfigure}
\caption{The distributions of the top twenty URL domains in each part of the ClueWeb22 dataset. The top twenty domains cover 12.6\% (B), 12.2\% (A), and 13.3 \% (L) of entire corpus. \label{fig:domain_dist}} 
\end{figure}

\section{Corpus Analysis}
\label{sec:stats}

This section analyzes the property of ClueWeb22, including the corpus distribution, statistics of document contents, and the quality of clean texts.

\subsection{Corpus Distribution}

We first analyze the distributions of URL domains, languages, and topic categories of ClueWeb22.

\textbf{URL Domain.} The distributions of top URL domains in ClueWeb22 are plotted in Figure~\ref{fig:domain_dist}. These top domains align with our intuitions about web popularity in search at 2022.

The top domains in ClueWeb22-B are those the search engine users around the world visit most frequently. They include high quality, trustworthy web sites such as Wikipedia and IMDB, as well as places people visit regularly for their daily activities, such as YouTube (entertainment) and Amazon (shopping). Popular websites from the big non-English speaking markets like China and Japan are also included. 
As discussed in previous sections, the documents are sampled through the view of one commercial search engine. The distribution reflects the perspective from the system. 
For example, the heavy focus of certain domains is perhaps a result of certain experiences offered by the search engine, for example question answering and domain-specific features.

Moving from ClueWeb22-B to ClueWeb22-A and then ClueWeb22-L, websites that cover a more diverse range of real-world activities become more popular. YouTube is the most popular site in ClueWeb22-A and ClueWeb22-L. Websites about jobs, food, and real estate are more popular. 

Compared to previous ClueWeb datasets, the top domains in ClueWeb22 reflect the dramatic changes of the internet in past decade. There is a significant growth of multi-media and user-created contents, often referred to as Web 2.0. 
Websites such as Quora, Yelp, and Instagram have become significantly more popular in the past decade.

\begin{figure}[t]
 \centering
          \begin{subfigure}[t]{0.25\textwidth}
         \centering
         \includegraphics[width=\textwidth]{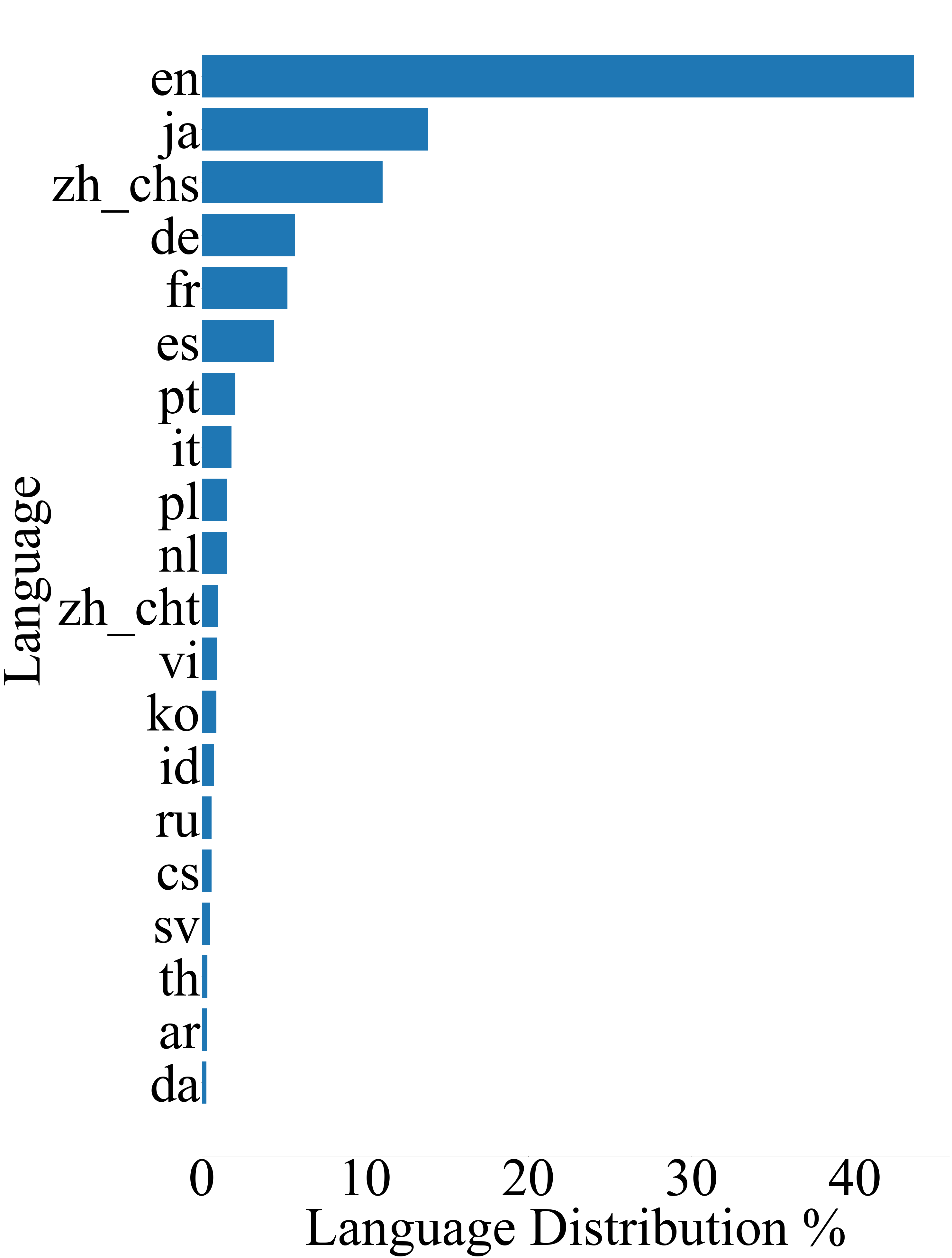}
         \caption{ClueWeb22-B.}
     \end{subfigure}
     \hspace{0.7cm}
     \begin{subfigure}[t]{0.25\textwidth}
         \centering
         \includegraphics[width=\textwidth]{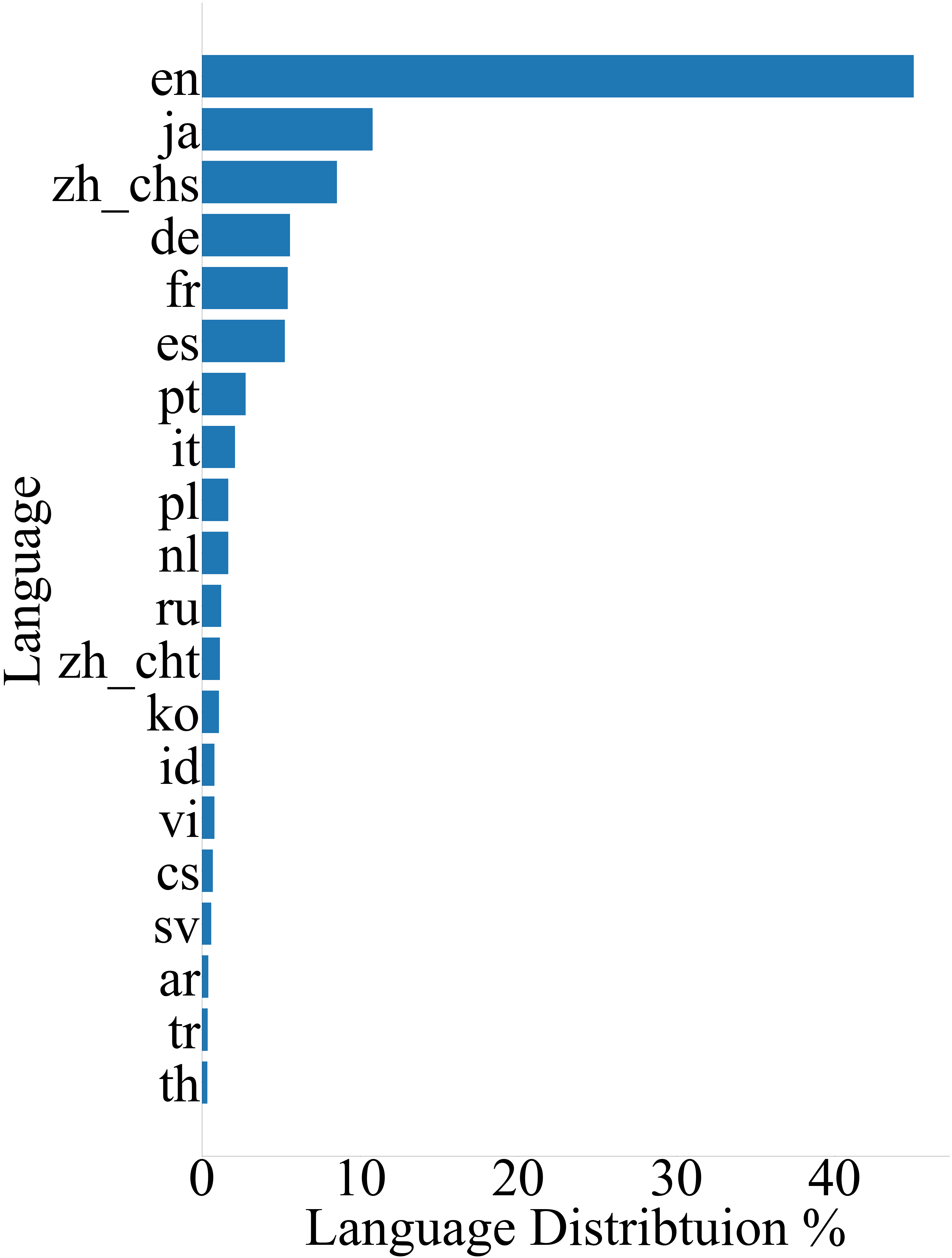}
         \caption{ClueWeb22-A.}
     \end{subfigure} \hspace{0.7cm}
          \begin{subfigure}[t]{0.25\textwidth}
         \centering
         \includegraphics[width=\textwidth]{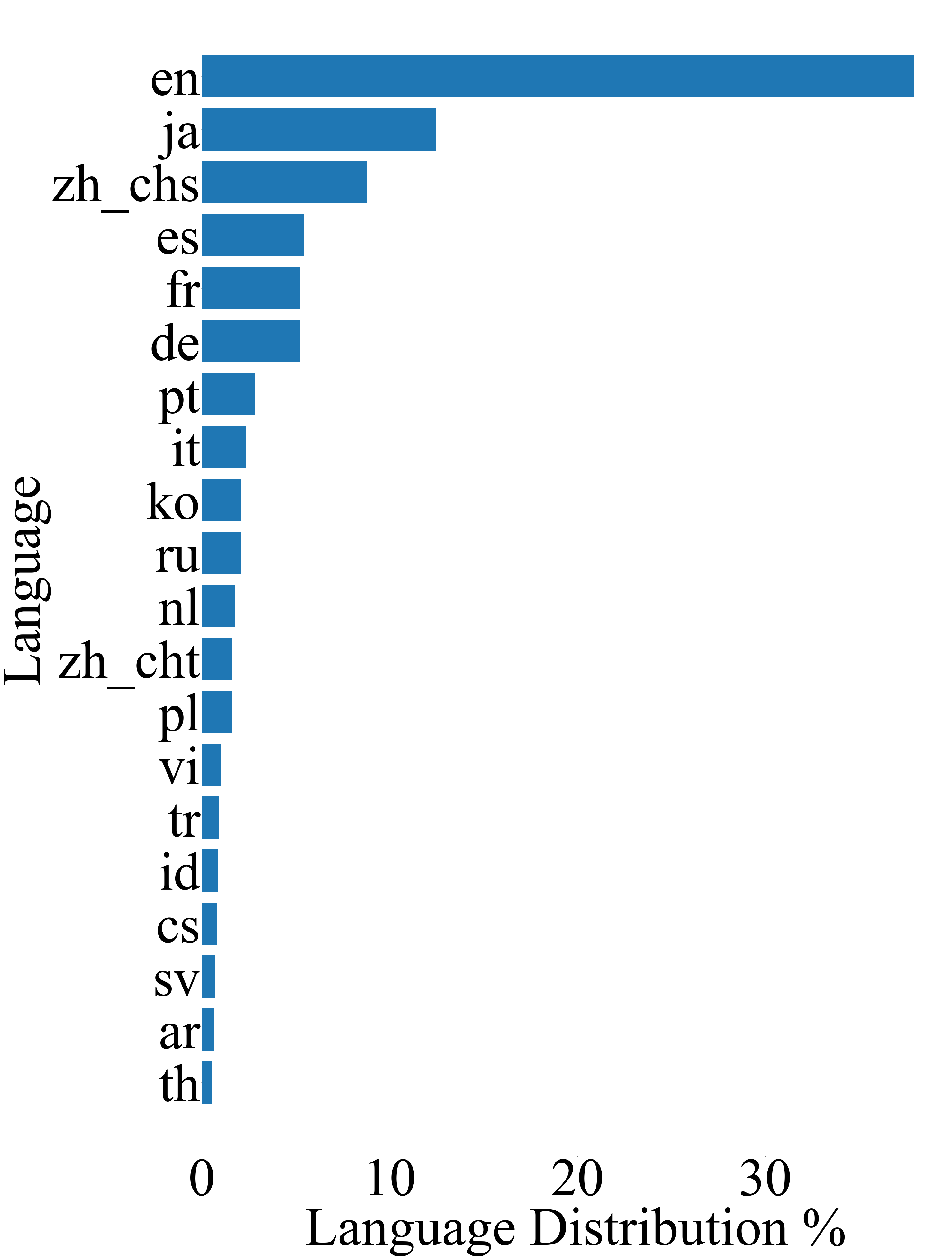}
         \caption{ClueWeb22-L.}
     \end{subfigure}
\caption{The distributions of the top twenty languages in ClueWeb22. 
The top twenty language cover 97.2\% (B), 96.3\% (A), and 95.0 \% (L) of the full set. \label{fig:language_dist}}
\end{figure}

\begin{figure}[t]
 \centering
          \begin{subfigure}[t]{0.3\textwidth}
         \centering
         \includegraphics[width=\textwidth]{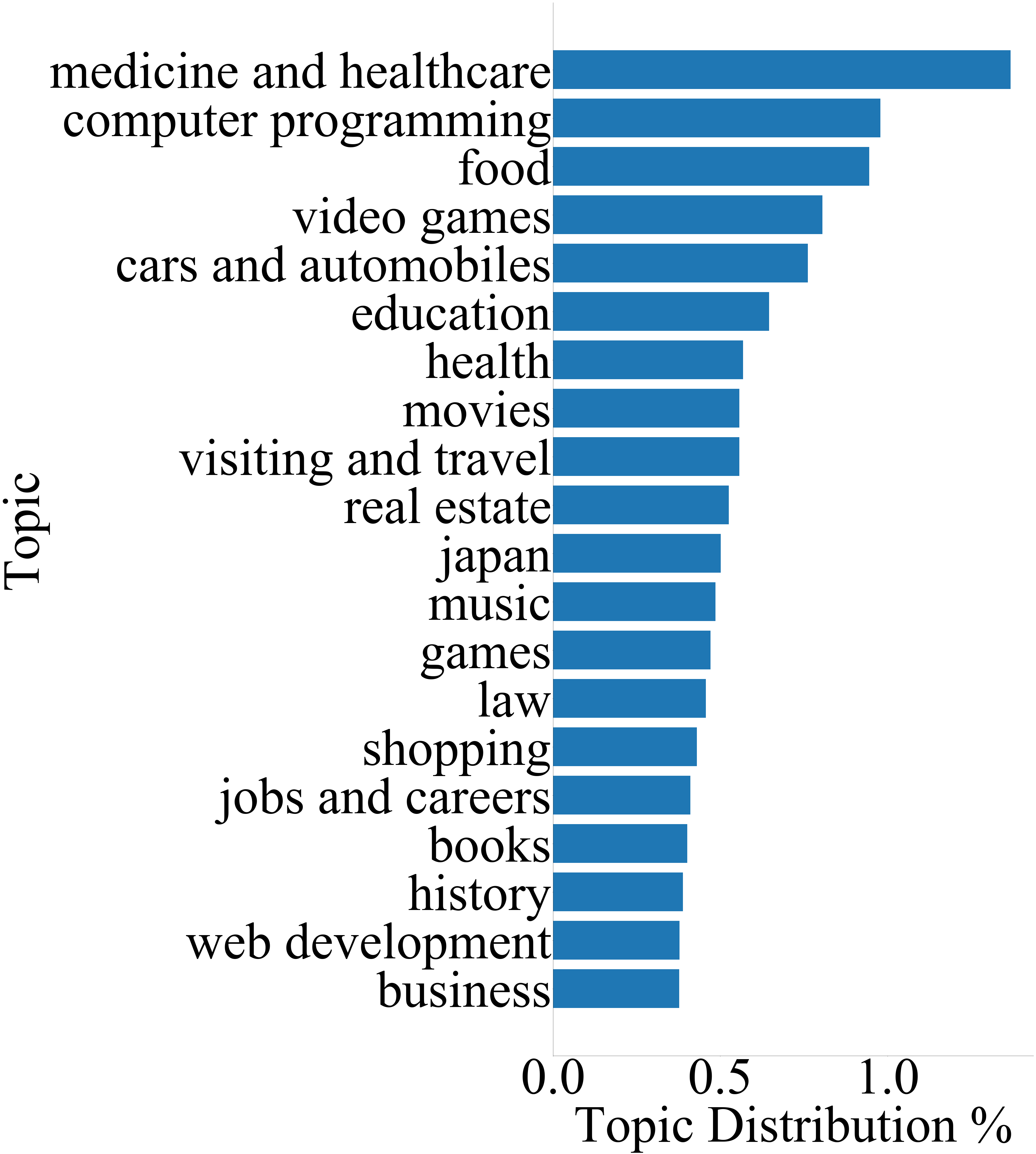}
         \caption{ClueWeb22-B.}
     \end{subfigure}
     \begin{subfigure}[t]{0.3\textwidth}
         \centering
         \includegraphics[width=\textwidth]{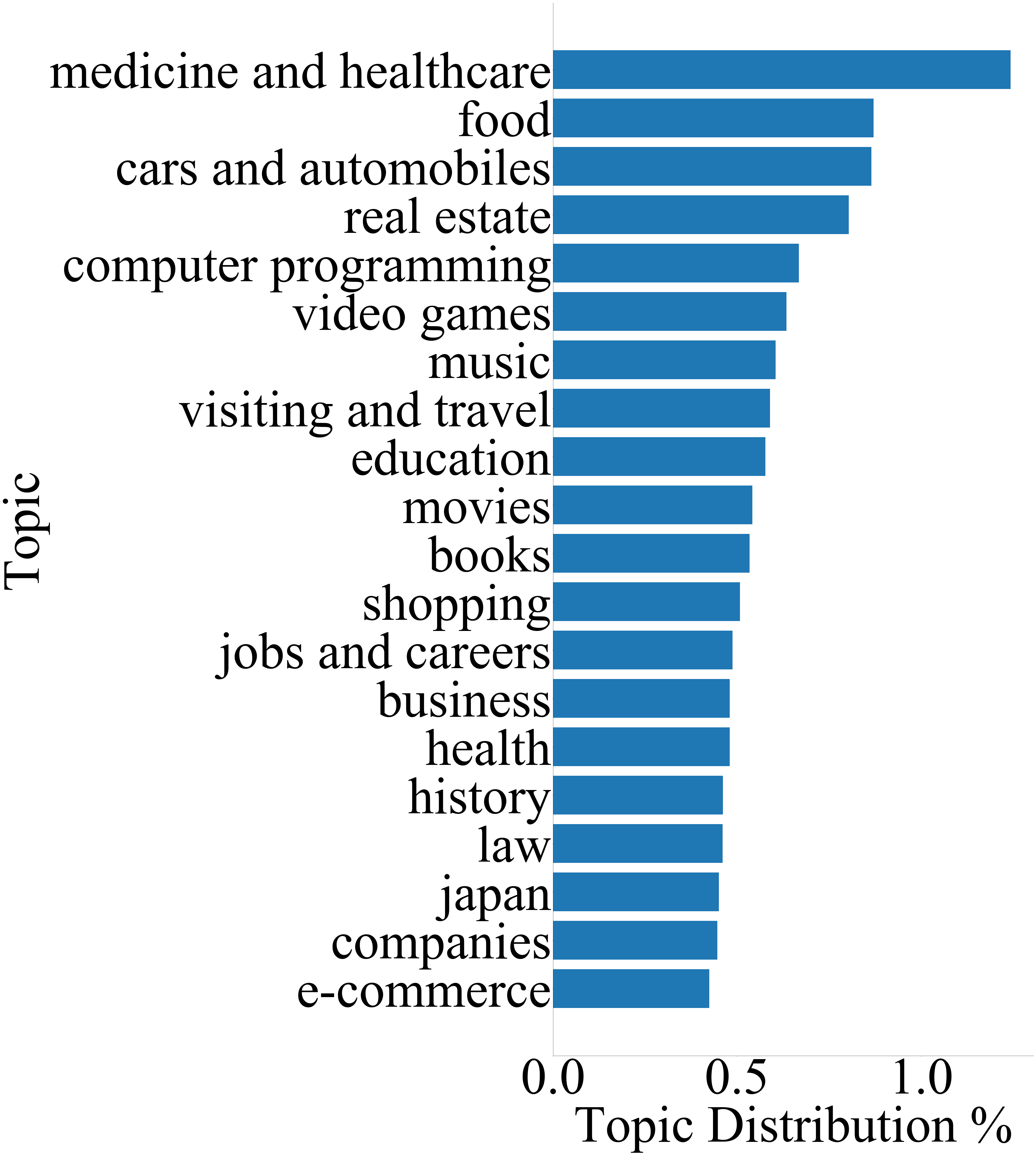}
         \caption{ClueWeb22-A.}
     \end{subfigure}
          \begin{subfigure}[t]{0.3\textwidth}
         \centering
         \includegraphics[width=\textwidth]{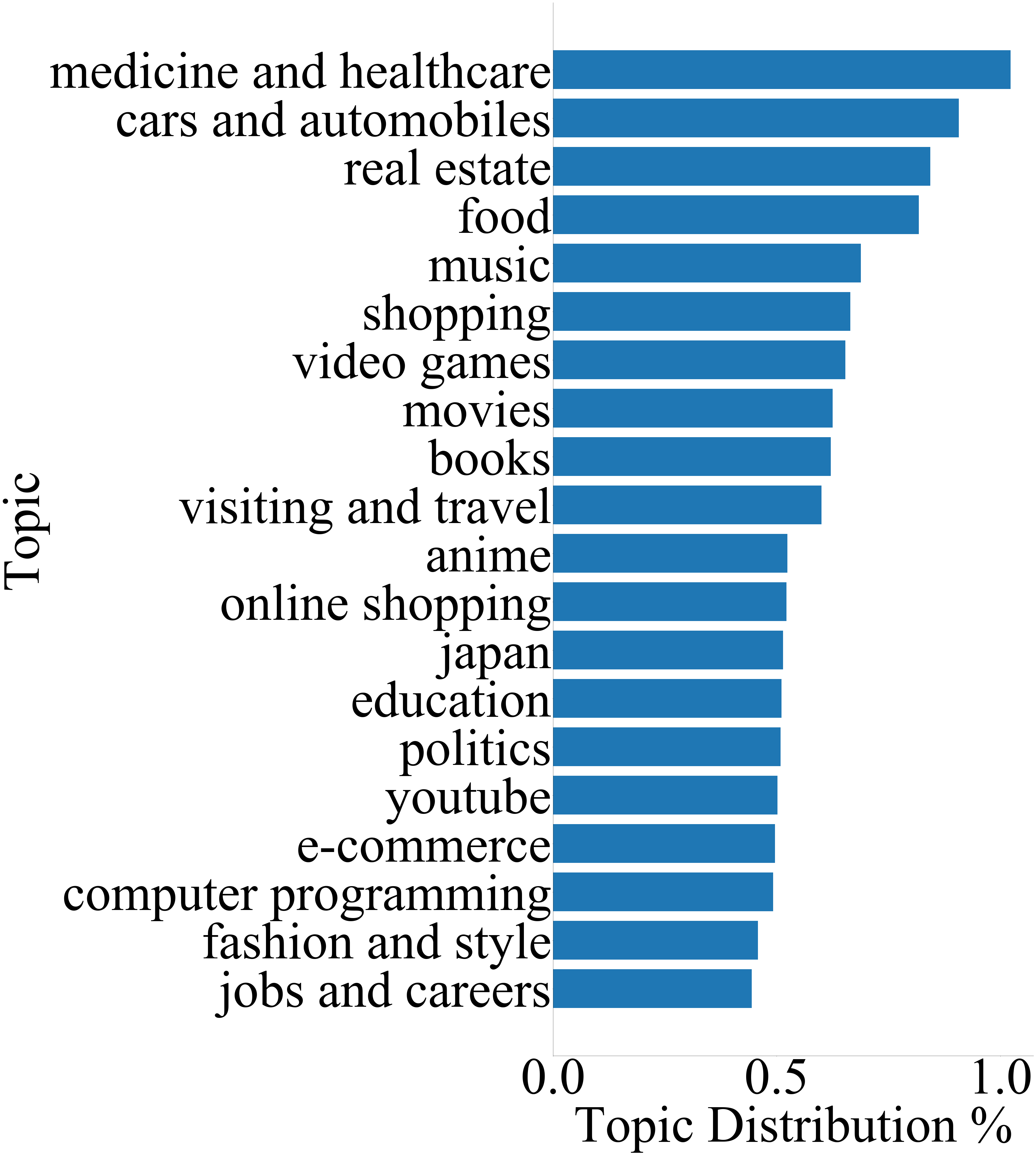}
         \caption{ClueWeb22-L.}
     \end{subfigure}
\caption{The distribution of the top twenty topics in ClueWeb22, classified by a standard text classifier. The top twenty topics cover  12.0\% (B), 12.1\% (A), and 12.7 \% (L) of the full set.} 
\label{fig:topic_dist}
\end{figure}

\textbf{Language.} Figure~\ref{fig:language_dist} plots the most popular languages in ClueWeb22. It is a direct count of the document language tags included in ClueWeb22 release. Note the language detection is done by an updated  version of BlingFire, which achieved 99\% detection accuracy for 365 languages~\citep{BlingFire}. 

As expected, English is the most frequent language in ClueWeb22. That said, the fraction of English documents is fewer than 50\% in ClueWeb22-B, and drops to below 40\% in ClueWeb22-L as the corpus becomes more diverse. The other top languages are correlated with the presence of the search engine in different markets, with strong appearances from eastern Asian in comparison to other web corpora.

The uneven language distribution is a known challenge for web data construction. Crawling web pages of one geographic location using machines in other places is difficult. The different characteristics of web pages in different markets also raise various challenges to navigate through the web. Although there is a long way to go to eliminate the biases towards certain languages, ClueWeb22 moves one step forward with more diverse coverage of non-English documents. In total ClueWeb22 includes 207 languages and has one of the highest fractions of non-English among publicly available web corpora.

\textbf{Topic.} In ClueWeb22 we include the predicted document topics as auxiliary data. The topic of a document is classified by a supervised neural model, using a standard network architecture.
It learns to classify documents into 100k topics. The topic catalog  is extracted from the web in a semi-automatic fashion. The training used weak supervision labels collected along the way. 
This weak supervision nature makes the classified topics more for analysis and reference purposes. We also like to note that the catalog and weak supervision labels are obtained from the western internet and thus many reflect certain focus and biases from there.

The top 20 topics are plotted in Figure~\ref{fig:topic_dist}. 
The popular topics reflect how users leverage search engines in their daily lives. 
Health related is the most popular among all categories. There are large appearances of transportation, food, accommodation, entertainment, and learning. 
There is also an emphasis on technology, especially on ClueWeb22-B.

\begin{figure}[t]
 \centering
          \begin{subfigure}[t]{0.3\textwidth}
         \centering
         \includegraphics[width=\textwidth]{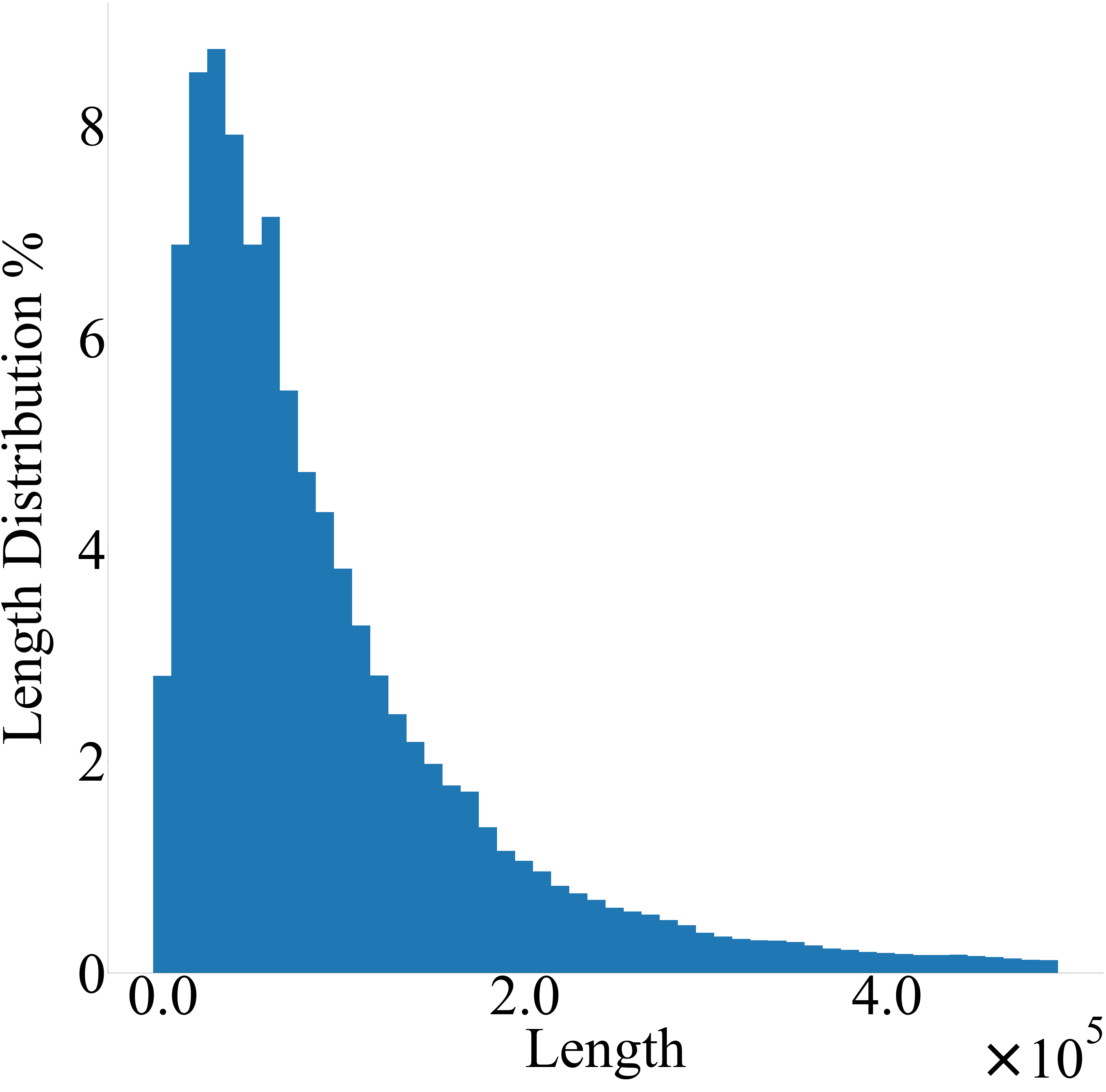}
         \caption{All HTML Character (117k)}
     \end{subfigure}
     \begin{subfigure}[t]{0.3\textwidth}
         \centering
         \includegraphics[width=\textwidth]{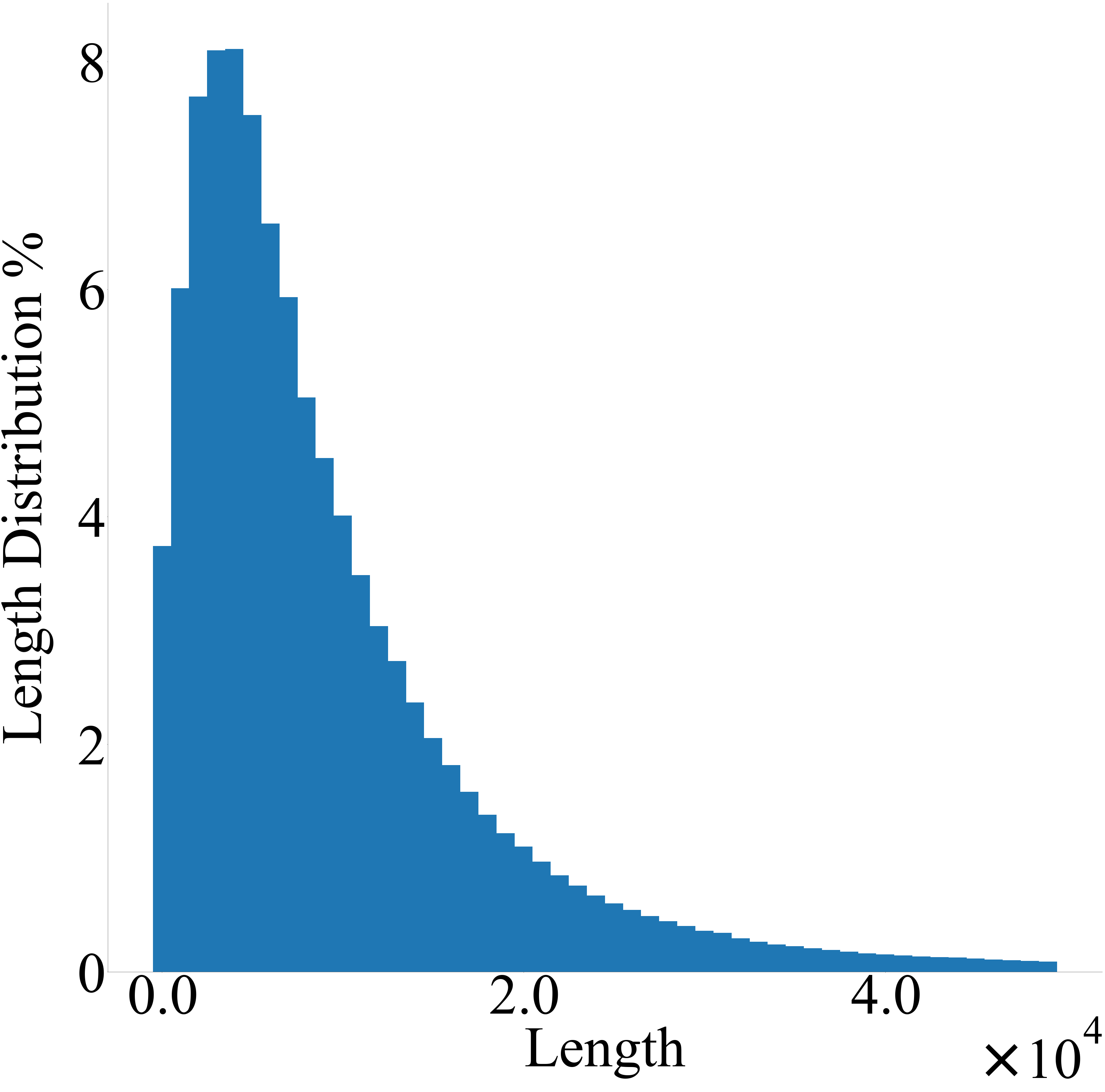}
         \caption{All Text Character (12k)}
     \end{subfigure}
          \begin{subfigure}[t]{0.3\textwidth}
         \centering
         \includegraphics[width=\textwidth]{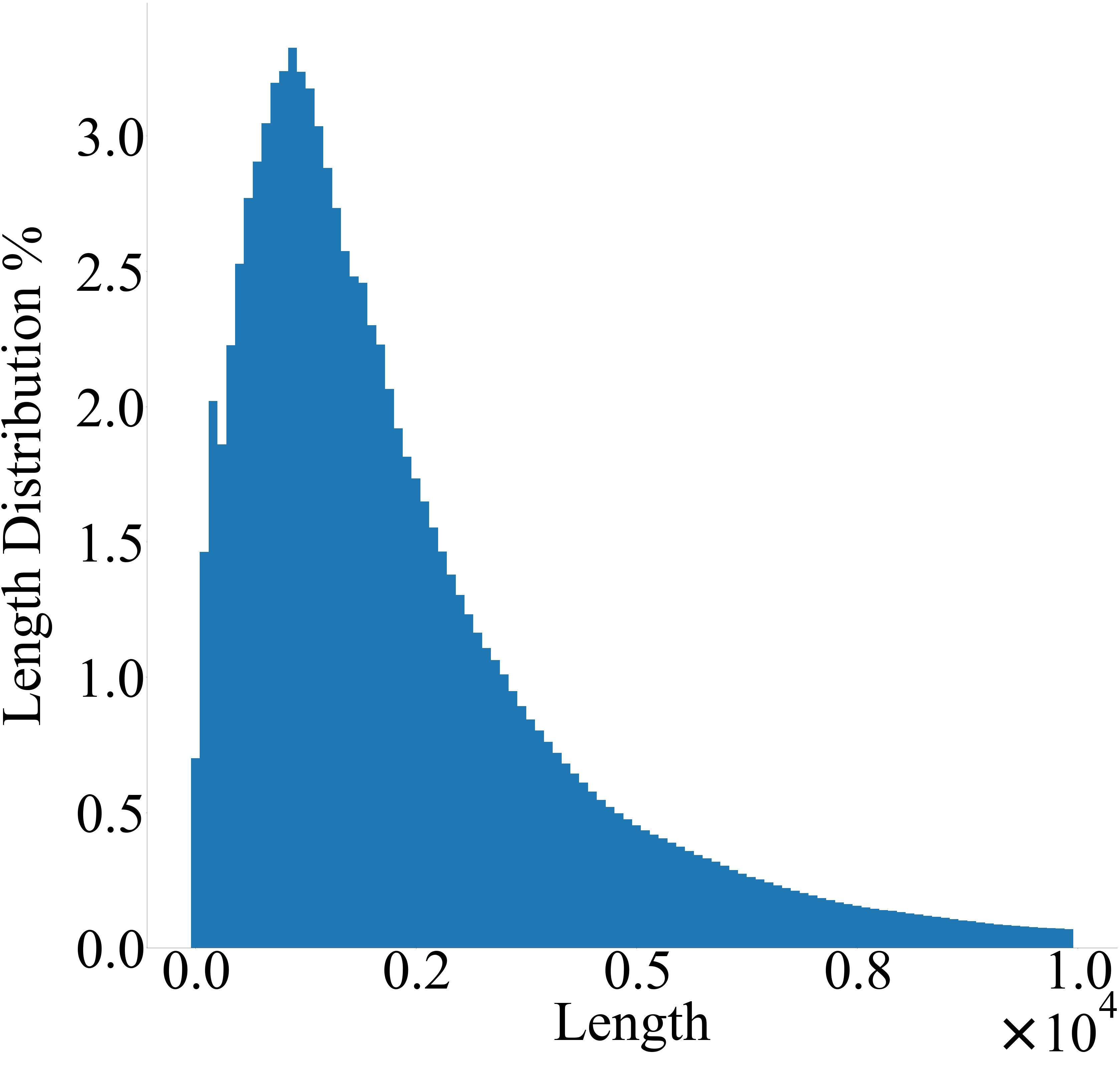}
         \caption{All Text Token (3k)}
     \end{subfigure}
\caption{The length distribution of HTML files of ClueWeb22  counted as the number of raw HTML characters (code and text),  characters of all texts, and tokens of all texts. The mean is shown in the parentheses.} 
\label{fig:html_len}
\end{figure}

\begin{figure}[t]
 \centering
          \begin{subfigure}[t]{0.3\textwidth}
         \centering
         \includegraphics[width=\textwidth]{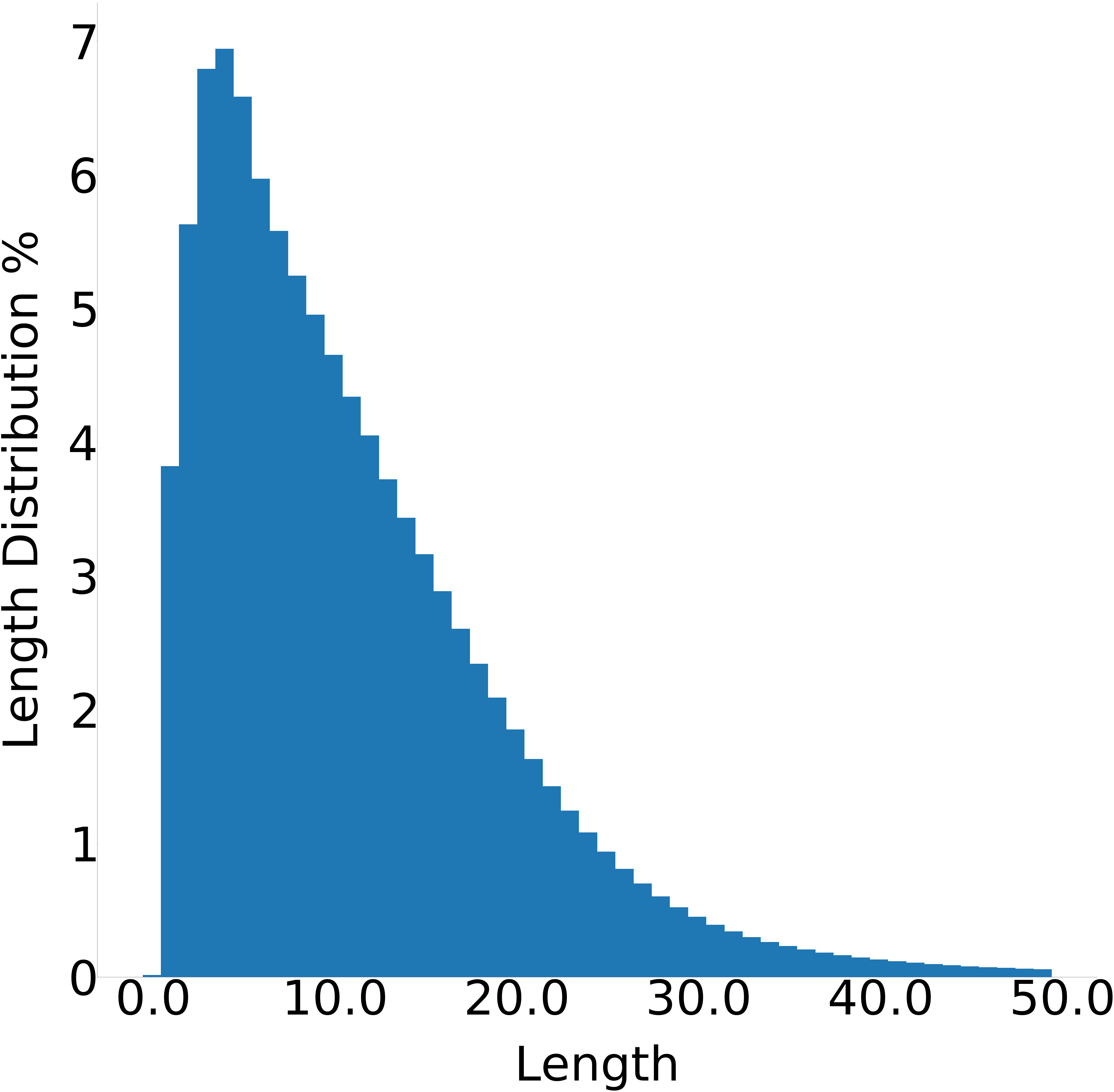}
         \caption{Title (13)}
     \end{subfigure}
     \begin{subfigure}[t]{0.3\textwidth}
         \centering
         \includegraphics[width=\textwidth]{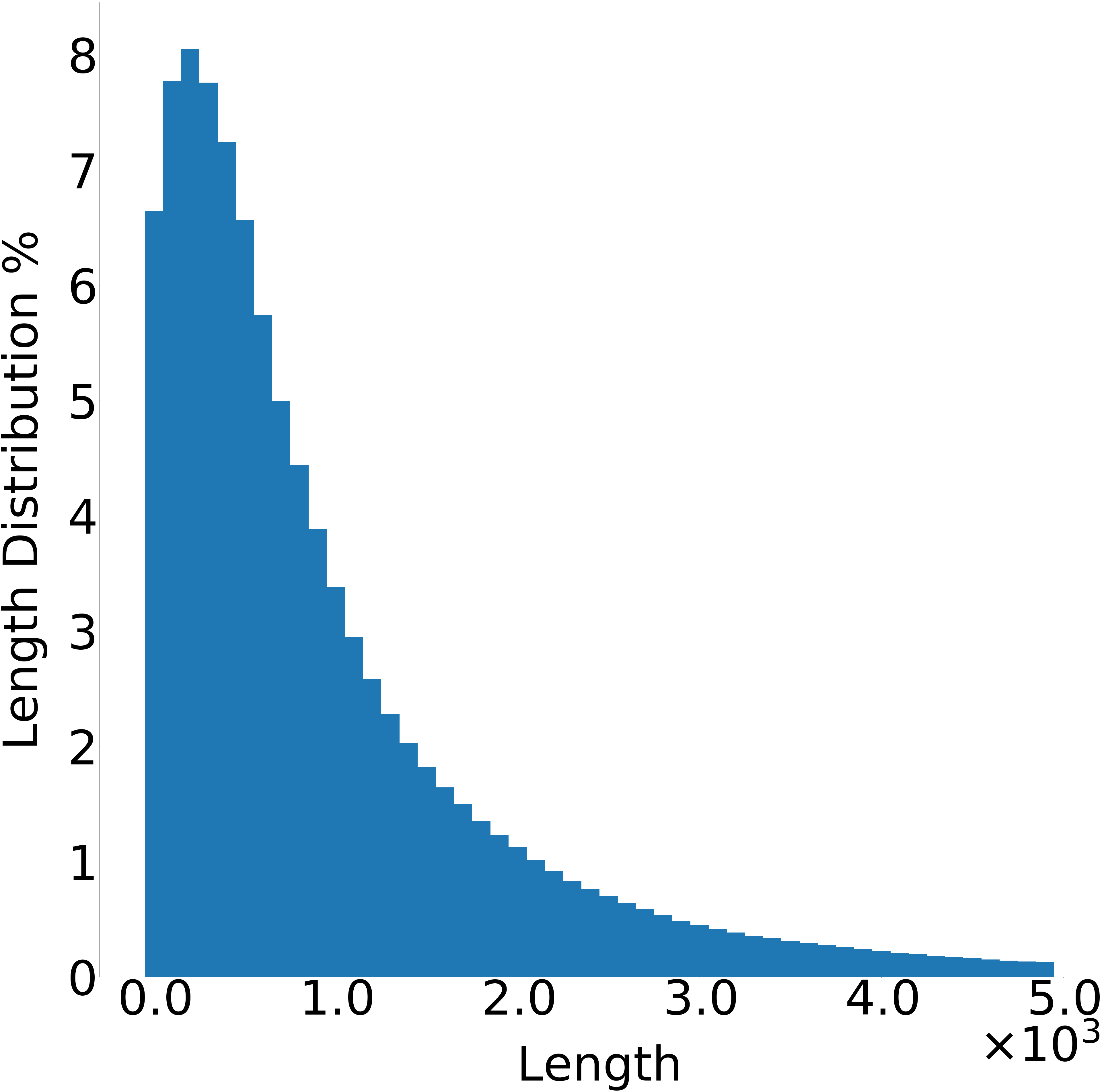}
         \caption{Primary Content (1.3k)}
     \end{subfigure}
          \begin{subfigure}[t]{0.3\textwidth}
         \centering
         \includegraphics[width=\textwidth]{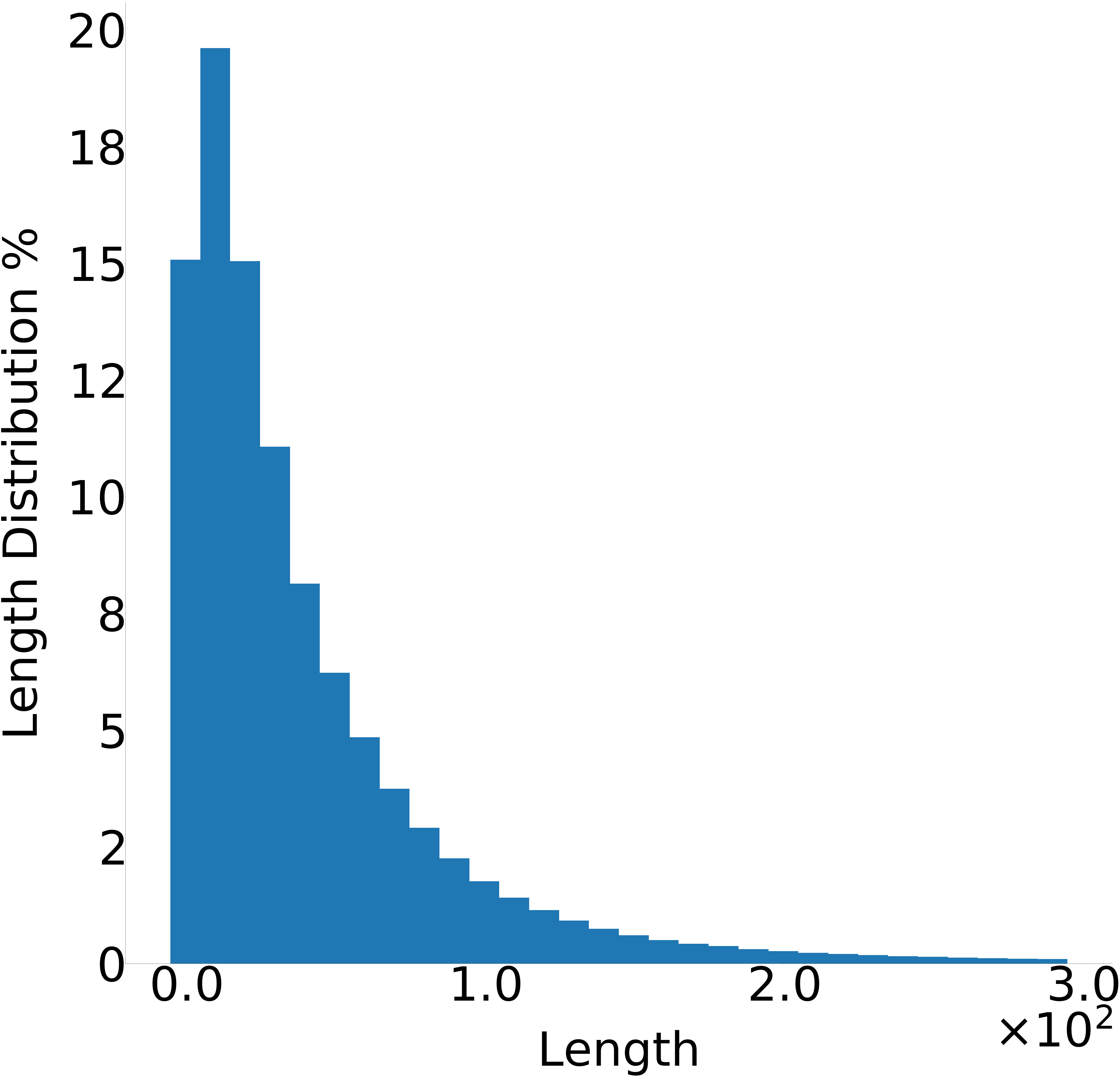}
         \caption{Paragraph (54)}
     \end{subfigure}
\caption{The length distributions of text tokens in title, primary content, and paragraphs in ClueWeb22, using XLM-R vocabulary~\citep{conneau2019unsupervised}. Average lengths are in parenthesis.
} 
\label{fig:text_len}
\end{figure}

\subsection{Document Content Statistics}
In the next part of this section, we conduct various studies on the extracted contents of ClueWeb22.


We first show the length distributions of ClueWeb22 pages, from the raw HTML characters, entire parsed texts, to the fine-grained document fields. All statistics are on the full ClueWeb22-L set when there is no significant difference between ClueWeb22 categorizes. 

\textbf{The length distributions of web pages} are plotted in Figure~\ref{fig:html_len}. We show three distributions.
\begin{enumerate}
    \item \texttt{RAW HTML Characters} are the original characters in the HTML of ClueWeb22 web pages, without any processing.
    \item \texttt{All Text Characters} are the characters of the directly extracted texts from the raw HTML, all fields mixed together, without any filtering and before our content extraction (which produces clean text latter). 
    \item \texttt{All Text Tokens} are the XLM-R~\citep{conneau2019unsupervised} sub-token counts from the directly extracted texts.
\end{enumerate}
These three statistics overview the raw data of ClueWeb22. The directly extracted text includes all text fragments in the HTML, many of which are noises and are the super set to be filtered for clean content.

The distributions are similar among three categories. They start with a Gaussian-like distribution on the shorter side, followed by a long tail distribution towards very long pages. 
Notably, the average number of text tokens per document is about 3k, with the 70th percentile at around 5k, much longer than the 512 maximum token length used in many pretrained language models.

\begin{table}[t]
    \centering
    \small
        \caption{The statistics of the number of semantic annotation tags per document in ClueWeb22. Primary content coverage is 100\% by definition, though a page can have empty primary content field.}
    \label{tab:annotation_count}
    \begin{tabular}{l|lll|lll|lll}
    \hline
    & \multicolumn{3}{c|}{\textbf{Total}} & \multicolumn{3}{c|}{\textbf{Average Per Doc}} & \multicolumn{3}{c}{\textbf{Standard Deviation}}  \\ 
     &  B & A & L & B & A & L  & B & A & L  \\ \hline
        Title & 153.6M&1.6B&8.1B & 0.8&0.8&0.8  & 0.6 & 0.6 &0.6 \\ 
        Heading & 2.3B&22.1B &  93.9B & 11.7 & 11.1 & 9.4 & 16.4 & 16.7 & 15.2\\ 
        Paragraph & 2.2B & 16.4B & 120.9B & 10.8 & 8.2 & 12.1 & 20.5 & 18.0 & 25.0 \\
        List & 273.3M & 2.0B & 6.9B & 1.4 & 1.0 & 0.7 & 4.5  & 3.7 &3.0 \\
        Table & 166.9M & 1.6B & 8.8B & 0.8 & 0.8 & 0.9 & 3.7 & 3.3 & 4.0 \\ 
         \hline         
    \end{tabular}
\end{table}

\textbf{The statistics of extracted document fields} are shown in Figure~\ref{fig:text_len}. We focus on three fields, title, primary content, and paragraph. The last one splits the primary content field.
The length of these extracted fields followed similar distribution trends with those at the raw HTML level. The average length of primary content is 1,352, about 44\% of the full text in Figure~\ref{fig:html_len}. Majority of raw texts in web pages are not primary and filtered. 

Content cleaning is a critical component in many web systems but often overlooked in the research community, partly due to lack of sufficient data. 
We hope the rich information released with ClueWeb22 can demonstrate the importance of better content extraction and facilitate more research in this under-explored direction.

\textbf{The statistics of structured annotations.}  We calculate the number of semantic annotation tags in ClueWeb22, e.g., the number of Tables detected in a web page.  Table~\ref{tab:annotation_count} lists the total, average, and standard deviation of annotated fields per document. 

Twenty percent of web pages do not have explicit HTML titles. Many web designers use other ways to present titles in pages, another reflection on the diverse nature of web page construction. 
The web is also quite structured. There are about 10 headings, 1 list, and 0.8 tables extracted per page. 
The list and table annotations in ClueWeb provide large scale structured information to be explored in directions such as information extraction, question answering, and structured data understanding.

\begin{table}[t]
    \centering
    \small
        \caption{The statistics of anchor links in ClueWeb22, including inlinks and outlinks. The numbers after removing hyperlinks from Header and Footers of web pages (w.o. H\&F) and after removing hyperlinks in the same web domain (w.o. In-Domain) are also listed. Average and standard deviation are the statistics per document for Total Number and the per anchor texts for Token Length.}

    \label{tab:anchor_stats}
    \begin{tabular}{l|rrr|rrr|rrr}
    \hline
    & \multicolumn{3}{c|}{\textbf{Total}} & \multicolumn{3}{c|}{\textbf{Average}} & \multicolumn{3}{c}{\textbf{Standard Deviation}}  \\ 
     &  B & A & L & B & A & L  & B & A & L  \\ \hline
        \textbf{Inlink} &&&&&&&&\\
         \ \textbf{Total Number} & 9.1B & 31.2B & 52.7B &  49.4  &  17.6  &  12.1  &  155.9  &  85.0  &  64.9 \\
         \  w.o. H\&F & 6.5B & 23.0B & 39.8B &  35.3  &  13.0  &  9.1  &  123.0  &  67.4  &  51.9 \\
         \  w.o. H\&F\&In-Domain & 548M & 1.1B & 1.5B &  3.0  &  0.6  &  0.3  &  29.3  &  12.8  &  9.0 \\ \hline
         \ \textbf{Token Length} & 75.5B & 263.6B & 452.4B &  8.3  &  8.5  &  8.6  &  132.2  &  107.8  &  92.1\\
         \  w.o. H\&F & 60.4B & 218.1B & 382.2B &  9.3  &  9.5  &  9.6  &  109.8  &  88.6  &  77.2\\
         \  w.o. H\&F\&In-Domain & 4.4B & 9.1B & 12.8B &  8.0  &  8.2  &  8.4  &  57.7  &  75.7  &  86.8\\
         \hline
         \textbf{Outlink}  &&&&&&&&\\
         \ \textbf{Total Number} & 22.3B & 211.5B & - & 112.1 & 106.1 & - & 125.9 & 121.1 & -\\
         \  w.o. H\&F & 16.1B & 151.6B & - & 81.0 & 76.0 & - & 109.9 & 104.6 & -\\
         \  w.o. H\&F\&In-Domain & 2.3B & 18.5B & - & 11.4 & 9.3 & - & 29.8 & 26.3 & - \\ \hline
         \ \textbf{Token Length} & 174.2B & 1.5T & - & 7.8 & 7.2 & - & 89.8 & 77.2 & -\\
         \  w.o. H\&F & 144.3B & 1.3T & - & 8.9 & 8.3 & - & 89.4 & 70.4 & -\\
         \  w.o. H\&F\&In-Domain & 18.6B & 130.6B & - & 8.2 & 7.1 & - & 133.6 & 97.2 & -\\
         \hline
    \end{tabular}
\end{table}
\begin{table}[t]
    \centering
    \small
        \caption{The fraction of outlink target in ClueWeb22 from ClueWeb22-B and ClueWeb22-A to ClueWeb22 categories and outside ClueWeb22. All anchors, those not appear in header and footer (w.o. H\&F) and not in-domain (w.o. In-Domain) are shown.     \label{tab:anchor_outlink}}

    \begin{tabular}{l|rrrr}
    \hline
   \textbf{From} & \textbf{To ClueWeb22-B} & \textbf{To ClueWeb22-A} & \textbf{To ClueWeb22-L} & \textbf{Outside}  \\ \hline
     ClueWeb22-B &24.1\% & 44.1\% & 54.9\% & 45.1\%\\
     \  w.o. H\&F &21.4\% & 40.3\% & 51.6\% & 48.4\%\\
     \  w.o. H\&F \&In-Domain & 12.9\% & 19.5\% & 22.5\% & 77.5\%\\
     \hline
    ClueWeb22-A &19.0\% & 39.4\% & 52.2\% & 47.8\%\\
     \  w.o. H\&F & 15.8\% & 35.0\% & 48.4\% & 51.6\%\\
     \  w.o. H\&F\&In-Domain &14.3\% & 21.2\% & 24.0\% & 76.0\%\\
     \hline
         
    \end{tabular}
\end{table}
\begin{table}[t]
    \centering
    \small
        \caption{The fraction of inlink source in ClueWeb22 before and after applying the without header footer (w.o. H\&F) and without in-domain link (w.o. In-D) filters. Note that all inlinks are extracted from the HTML pages of ClueWeb22-A. \label{tab:anchor_inlink}}

    \resizebox{\textwidth}{!}{
    \begin{tabular}{l|lll|lll|lll}
    \hline
  \textbf{To} $\rightarrow$  & \multicolumn{3}{c|}{\textbf{ClueWeb22-B}} & \multicolumn{3}{c|}{\textbf{ClueWeb22-A}} & \multicolumn{3}{c}{\textbf{ClueWeb22-L}}  \\ \hline
  \multirow{2}{*}{\textbf{From} $\downarrow$}   &  \multirow{2}{*}{Total} & \multirow{2}{*}{w.o. H\&F} & w.o. H\&F &  \multirow{2}{*}{Total} & \multirow{2}{*}{w.o. H\&F} & w.o. H\&F   &  \multirow{2}{*}{Total} & \multirow{2}{*}{w.o. H\&F} & w.o. H\&F \\ 
  & & & \&In-Domain   & & & \&In-Domain  & & & \&In-Domain 
  
  \\
  \hline
      From ClueWeb22-B & 26.3\% & 26.9\%& 18.9\% & 17.1\% & 16.9\% & 17.4\% & 14.0\% & 13.8\% & 16.6\%\\
      From ClueWeb22-A $\setminus$ B & 73.7\% & 73.1\% & 81.1\% & 86.8\% & 87.1\% & 84.1\% & 90.3\% & 90.4\% & 85.4\%\\
         \hline
    \end{tabular}}
\end{table}

\textbf{Anchor Graph.} The statistics of anchor graphs are shown in Table~\ref{tab:anchor_stats}. We show the statistics of the raw anchor graph data and that after two filtering strategies. The first excludes hyperlinks in headers and footers, using the flags included in the anchor data. The second  filters out hyperlinks that connect pages from the same domain, in addition to removing header and footer links. 

In total there is a large amount of anchor links extracted from ClueWeb22-A. Most of them are in-domain. The number of inlinks per page drops significantly from ClueWeb22-B to ClueWeb22-A and then to ClueWeb22-L. Head web pages received more inlinks than tail ones as expected.
The numbers of outlinks per page in ClueWeb22-B and ClueWeb22-A are relatively the same.

The drastic drop of anchor links after removing in-domain ones shows that most hyperlinks are pointing to other pages from the same website. How to better identify anchor links that are more informative, i.e., including semantic information of the destination URLs is an important research question to make best use of the anchor information.

We also provide a preliminary analysis on the connectivity of the anchor graph in ClueWeb22. The fraction of outlinks destination is shown in Table~\ref{tab:anchor_outlink}. The source of inlinks is in Table~\ref{tab:anchor_inlink}. After the two filters, more than 75\% of outlinks in ClueWeb22-B and ClueWeb22-A point to web pages outside of ClueWeb22. This shows the enormous nature of the web. ClueWeb22-B received more hyperlinks from both ClueWeb22-B and ClueWeb22-A, 13\% and 14\%, in comparison to its size, which is 200 Million out of 10 billion (2\%). This indicates that ClueWeb22-B is sampled from web sites closer to the center and more connected part of the web graph. 
The ClueWeb22-B subset, which is 20\% of ClueWeb22-A, provided slightly less than 20\% of inlinks, on the other hand.

\subsection{Clean Text Quality}

Manual examination indicates that the quality of ClueWeb22 clean text, especially in ClueWeb22-B and ClueWeb22-A, is pretty good. They are sampled from the popular part of the web, filtered by production quality systems, and extracted by advanced content understanding techniques. 

To demonstrate the content quality of ClueWeb22 documents in a quantitative way, we conduct a pilot study using an important application of large web corpora: to pretrain language models. 
As a preliminary study, we choose the most common pretraining and downstream pipeline: pretrain vanilla RoBERTa base~\citep{liu2019roberta} and evaluate by fine-tuning on the MNLI task~\citep{MNLI}.

This study used a code base similar to the open-source version of Efficient LM Pretraining.\footnote{https://github.com/microsoft/Efficient-Large-LM-Trainer} We followed the exact pretraining and fine-tuning configurations of the RoBERTa base baseline by a previous research~\citep{meng2021coco}. We used the \textit{base} configuration which is to pretrain the 12 layer BERT for 125k steps, with a maximum 512 tokens per sequence and the global batch size of 2048 sequences. Everything was kept the same, except the pretraining corpus was changed from Wikipedia and Book Corpus to Wikipedia+ClueWeb22, using the  primary content of ClueWeb22-B documents.

Specifically, experiments were run with the following text resources, all in English.
\begin{enumerate}
    \item Wiki: the official Wikipedia English dump widely used for language model pretraining.
    \item Book: Google Book Corpus~\citep{zhu2015aligning}, which is often paired with Wiki to form the \textit{base} pretraining setup.
    \item ClueWeb22-B Random: the primary content from randomly sampled ClueWeb22-B web pages.
\end{enumerate}

\begin{table}[t]
    \centering
    \small
      \caption{MNLI Dev accuracy of RoBERTa$_\text{base}$ model pretrained on different combinations of ClueWeb22-B documents, Wikipedia, and Book Corpus. The pretraining and fine-tuning use the exact same setting, following the standard configurations in recent research~\citep{meng2021coco}. Only pretraining text corpora differ.  \label{tab:glue}}
    \begin{tabular}{l|l|l|l}
    \hline
    \textbf{Model} & \textbf{Batch $\times$ Step} & \textbf{Pretraining Corpus (size)} & \textbf{MNLI-m/mm} \\ \hline
 RoBERTa$_\text{base}$ & 2K$\times$125K & Wiki (12GB) + Book (6GB) & 86.2/85.9 \\
 RoBERTa$_\text{base}$ & 2K$\times$125K & Wiki (12GB) + ClueWeb22-B Random (6GB) & 85.8/86.2 \\
 RoBERTa$_\text{base}$ & 2K$\times$125K & Wiki (12GB) + ClueWeb22-B Random (18GB) & 86.2/86.4 \\
    \hline
    \end{tabular}
\end{table}

The MNLI accuracy of the same RoBERTa$_\text{base}$ model pretrained on different corpora is listed in Table~\ref{tab:glue}.
Our RoBERTa$_\text{base}$ results with Wiki+Book  are on par or better than those reported in previous research~\citep{liu2019roberta, meng2021coco}.
Pretraining with 6 GB of randomly-selected ClueWeb22-B web pages, in replacement of the Book corpus and combined with Wiki, leads to similar MNLI accuracy. This shows the quality of ClueWeb22-B primary content is on par with the widely used Book Corpus for language model pretraining.
Increasing the amount of ClueWeb22-B text to 18GB further improved MNLI accuracy, indicating the potential of using more high quality texts from ClueWeb22 for pretraining. 

In comparison, previous research~\citep{raffel2019t5} found that the C4 web corpus derived from CommonCrawl, even after strong filters, does not provide pretraining texts as high quality as Wiki+Book. The advantage of CommonCrawl is at its scale but not necessary content quality.
Our pilot study shows that ClueWeb22-B provides a large amount of high quality text that can be directly used (without any further cleaning) to pretrain language models.

\section{Comparison with CommonCrawl}
\label{sec:cc}

\begin{table}[t]
    \centering
    \small
        \caption{Statistics of July and August 2022 CommonCrawl (CC) Snapshots. Distinct pages are de-duplicated on CommonCrawl URLs. Success Crawl only keep those pages with ``HTTP 200'' responses which indicate the crawling was successful. The de-duplication and successful crawl filtering are applied when constructing ClueWeb22.}
    \label{tab:ccdoc}
    \begin{tabular}{lrrr|rr}
    \toprule
        & \multicolumn{3}{c|}{\textbf{Page Level}}  & \multicolumn{2}{c}{\textbf{Domain Level}}\\
    & {\#All}  & {\#Distinct} & {\#Success Crawl}  & {\#All} & {\#Success Crawl}\\ \hline
        July 2022 CommonCrawl &  3.6B & 3.6B & 3.1B & 40.8M & 37.2M\\ 
        August 2022 CommonCrawl &  3.3B & 3.2B & 2.6B & 43.0M & 38.7M\\ 
        ClueWeb22-B & 200M & 200M & 200M & 15.0M & 15.0M\\ 
        ClueWeb22-A & 2B & 2B & 2B & 45.3M & 45.3M\\
        ClueWeb22-L & 10B & 10B & 10B & 66.0M & 66.0M\\
        \hline
        \textbf{Overlap} & & {\#Distinct} & {\#Success Crawl}  & {\#All} & {\#Success Crawl}\\ \hline
        \multicolumn{2}{l}{July 2022 CommonCrawl vs. August 2022} & 126.5M & 104.3M& 33.9M & 31.5M \\
        \multicolumn{2}{l}{July 2022 CommonCrawl vs. ClueWeb22-B} & 28.5M & 27.6M & 11.2M & 10.6M\\
        \multicolumn{2}{l}{July 2022 CommonCrawl vs. ClueWeb22-A} & 145.4M & 141.3M & 24.0M & 22.7M\\
        \multicolumn{2}{l}{July 2022 CommonCrawl vs. ClueWeb22-L} & 454.9M & 444.3M & 29.1M & 27.4M\\
        \multicolumn{2}{l}{August 2022 CommonCrawl vs. ClueWeb22-B} & 21.0M & 19.9M &11.4M &10.7M \\
        \multicolumn{2}{l}{August 2022 CommonCrawl vs. ClueWeb22-A} & 102.7M & 96.8M &24.6M &23.2M  \\
       \multicolumn{2}{l}{ August 2022 CommonCrawl vs. ClueWeb22-L} & 313.0M & 296.2M &30.0M &28.1M \\
         \bottomrule
    \end{tabular}
\end{table}

In this section we analyze the differences between ClueWeb22 and CommonCrawl. Their main differences are derived from different design choices in their construction. ClueWeb22 documents are samples of important web pages from a trillion-scale web crawl of a commercial search engine. CommonCrawl includes pages from a community maintained crawling process, at the scale of billions of pages, with a crawled snapshot produced every month.

\textbf{General CommonCrawl Statistics.} Specifically, we analyzed two CommonCrawl monthly snapshots, July 2022 and August 2022, for comparison with ClueWeb22, which was collected around February 2022. For the analysis, we perform basic de-duplication at URL level and also check the HTTP status of crawled web pages, where ``HTTP 200'' responses indicate the crawling of corresponding web pages were successful.

\begin{figure}[t]
 \centering
          \begin{subfigure}[t]{0.24\textwidth}
         \centering
         \includegraphics[width=\textwidth]{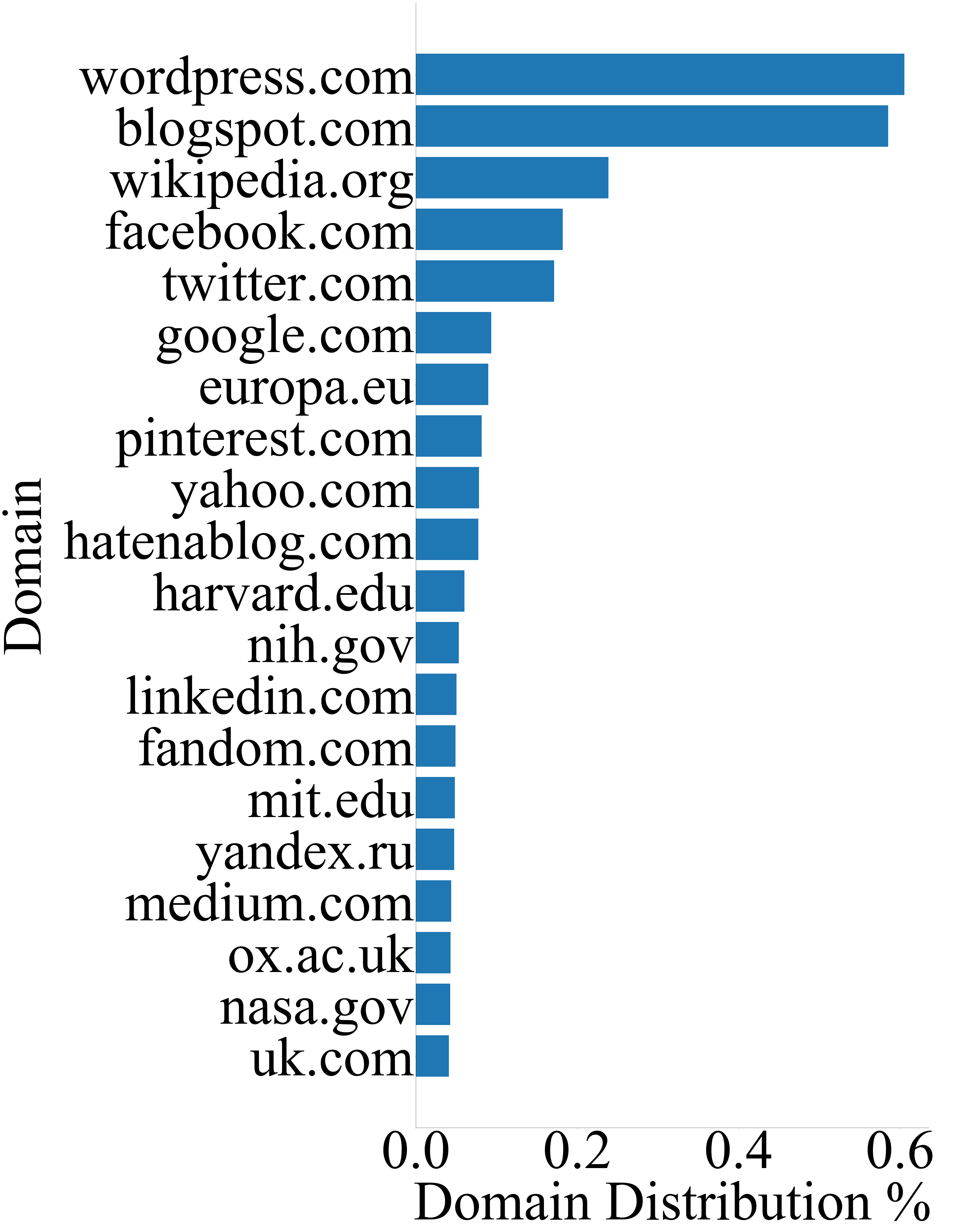}
         \caption{July 2022.}
     \end{subfigure}
     \begin{subfigure}[t]{0.24\textwidth}
         \centering
         \includegraphics[width=\textwidth]{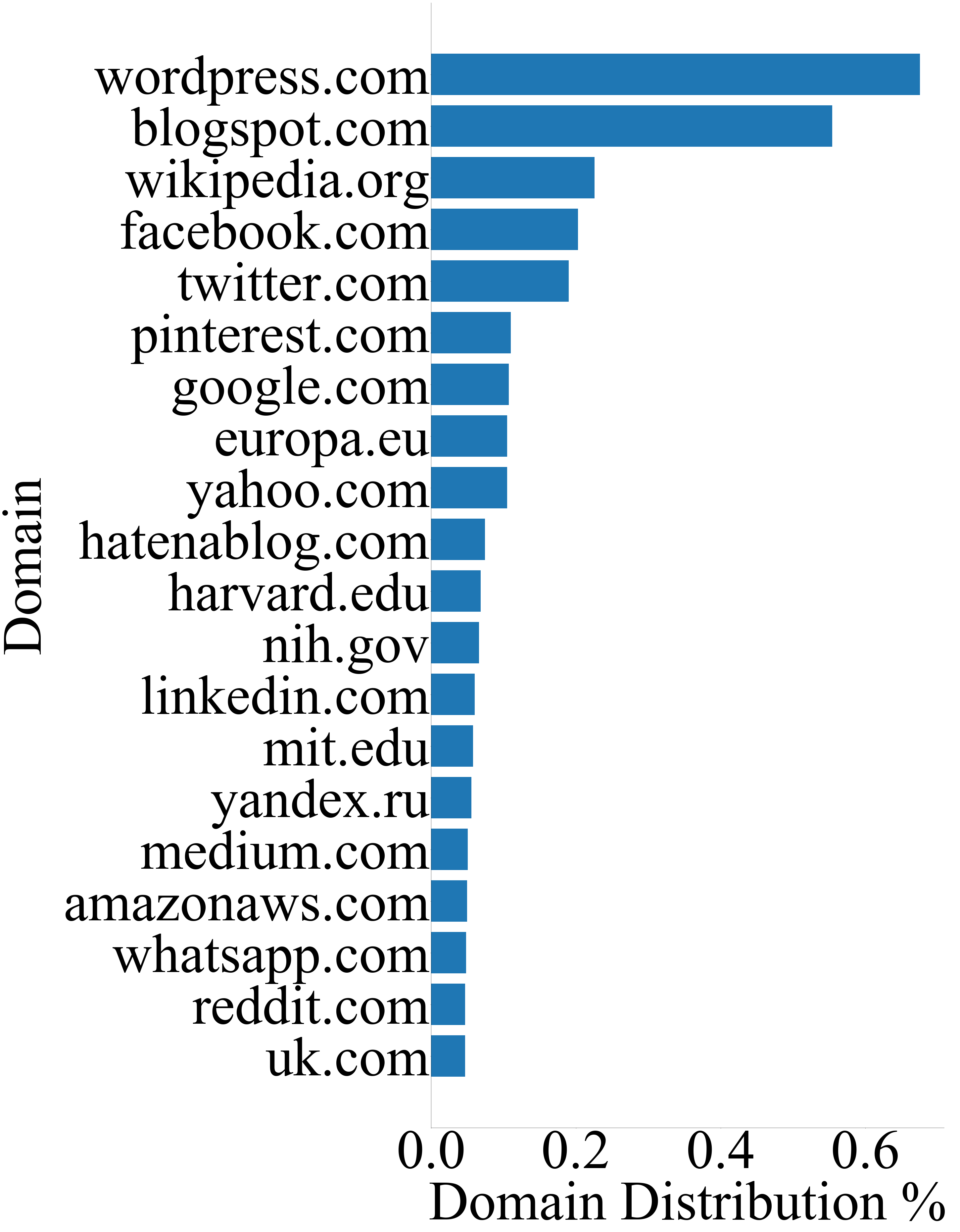}
         \caption{August 2022.}
     \end{subfigure}
     \begin{subfigure}[t]{0.24\textwidth}
         \centering
         \includegraphics[width=\textwidth]{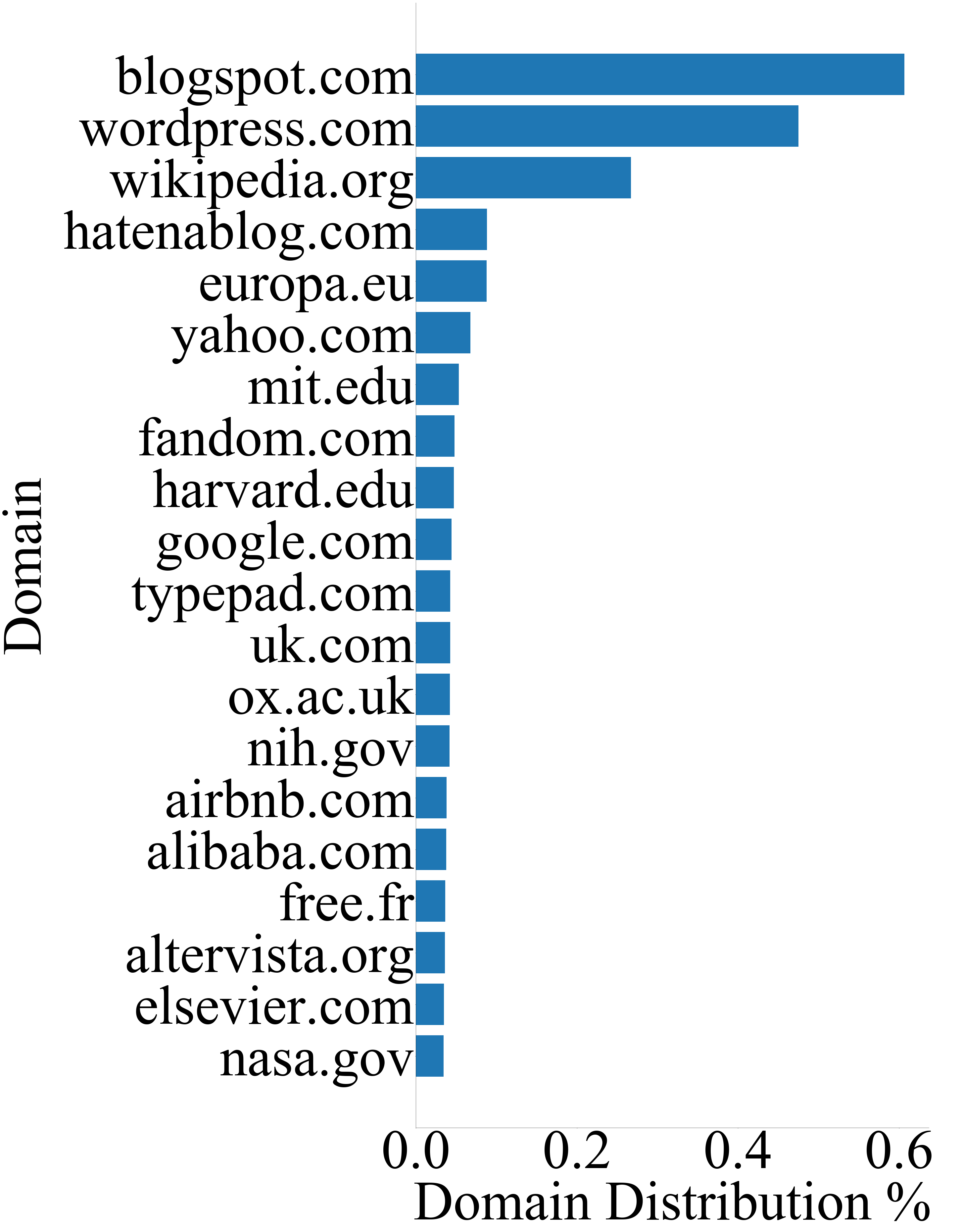}
         \caption{July 2022 Success Crawl.}
     \end{subfigure}
     \begin{subfigure}[t]{0.24\textwidth}
         \centering
         \includegraphics[width=\textwidth]{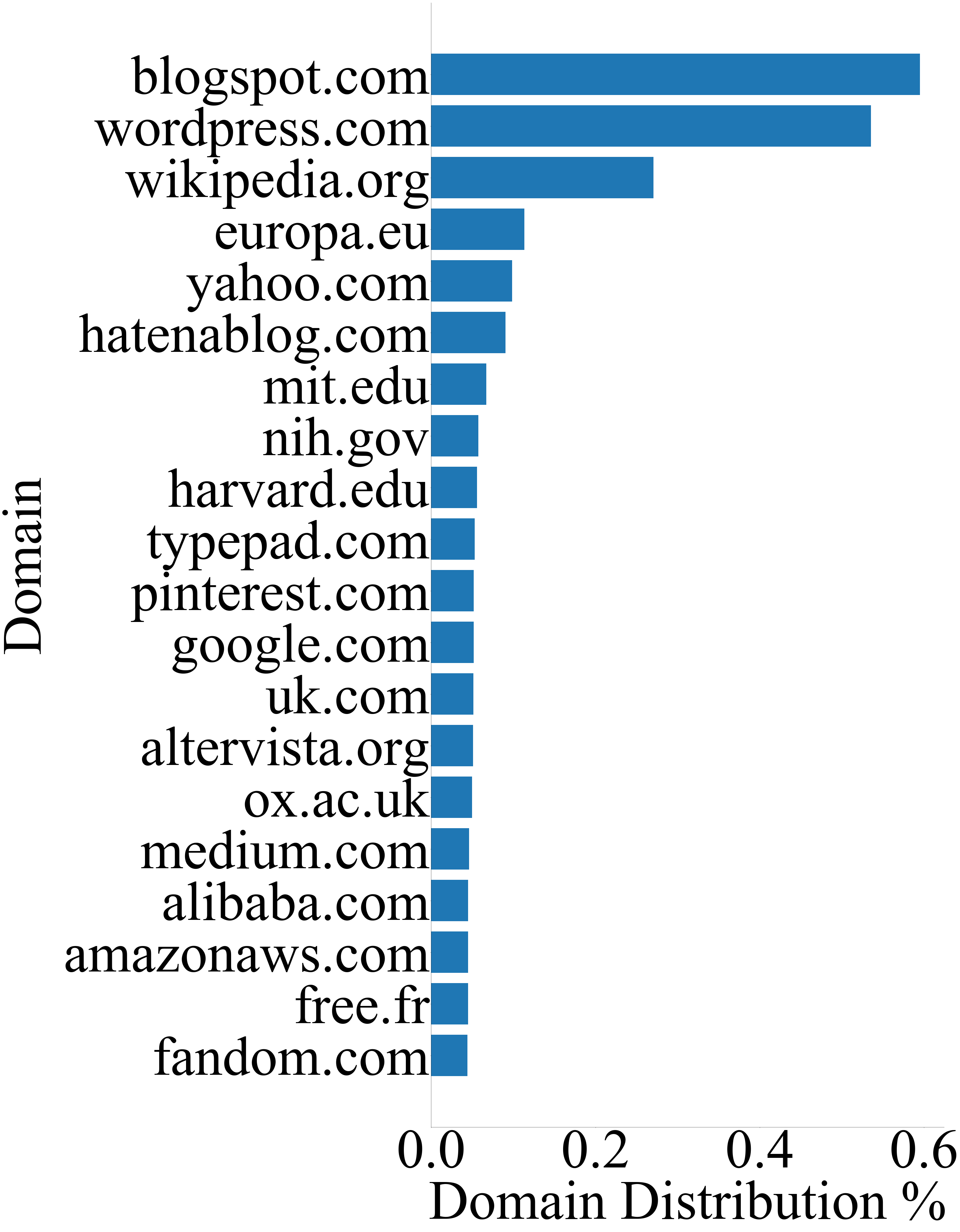}
         \caption{August 2022 Success Crawl.}
     \end{subfigure}
\caption{The distributions of the top twenty URL domains in CommonCrawl July and August 2022 snapshots. 
Success Crawls are pages with HTTP 200 responses. 
The top twenty domains cover around 2.5\% in each distribution.
} 
\label{fig:domain_dist_cc} 
\end{figure}

Table~\ref{tab:ccdoc} shows the number of URLs and web domains of these two snapshots. We also show these statistics of ClueWeb22 and the overlap between these datasets. 
The top domains of the two months before and after HTTP 200 filtering are plotted in Figure \ref{fig:domain_dist_cc}.  
These statistics show notable differences between CommonCrawl and ClueWeb22 in their coverage and distribution. We discuss them in the remainder of this section.

\textbf{Differences from Crawling.}  Each CommonCrawl snapshot is an individual crawl. There is little overlap between the URLs in the two CommonCrawl snapshots. This is beneficial because a large amount of unique URLs are accumulated throughout the history of CommonCrawl. They also differ significantly from the URLs included in ClueWeb22.
On the other hand, at the domain level, there are large overlaps between the domains included in the two CommonCrawl snapshots. It indicates that although the crawlers traversed different URLs, they stay in the similar part of the web. 
The high domain-level overlap suggests that the monthly CommonCrawl snapshots were samplings from similar distributions. The low page-level overlap indicates  that it is a sparse sample.

ClueWeb22 covers a different part of the web and has lower overlap with both CommonCrawl snapshots. ClueWeb22 pages are also spread more widely in different domains.
For example, in the July 2022 CommonCrawl there are 3.1 billion pages sampled from 37 million domains, 83 pages per domain on average. In comparison, ClueWeb22-A includes 2 billion pages from 45 million domains, averaging 44 pages/domain.

We also notice a significant drop on some top domains in CommonCrawl after the HTTP 200 filter is applied, for example, Google, twitter, and Pinterest. Many websites do not allow a large amount of requests from the same IP address and often put a frequency limit for crawlers they explicitly allow. 
There is more incentive for websites to allow crawling from a commercial search engine, for example, to be indexed and exposed to search users. ClueWeb22 benefits from this with better crawler reach in certain websites.

These differences are direct results of the crawling process. CommonCrawl provides more diversity along the time axis as it crawles and releases snapshots every month. ClueWeb22 provides more diversity at the domain level as it is sampled from the index of a commercial search engine who has a much deeper exploration of the web.

\textbf{Differences from Sampling.} CommonCrawl web pages are crawled based on ranking using their in-house web graphs, prioritizing discovered URLs with high harmonic centrality and PageRank~\citep{CCSampling}. This leads a standard link-structure based distribution as reflected in Figure~\ref{fig:domain_dist_cc}. 
Since Summer 2021, a per-domain limit has been enforced in CommonCrawl, with maximum 25 million URLs allowed per top domain and 150k URLs per host~\citep{CCSampling}. This rule is more restrictive for site with fewer hosts, as a site with one host would hit the per host budget before reaching the full 25 million quota.  This perhaps leads to the high fraction of wordpress.com and blogspot.com who have many hosts.

In comparison, ClueWeb22 pages are sampled using signals beyond link structures to estimate page importance. Many of the importance signals, especially those stemmed from user search behaviors, are quite powerful but not available for CommonCrawl.

There is no single best definition of web page importance. Nevertheless, 
the frequency of websites in CommonCrawl is unlikely to be aligned with their frequency of being visited by web search users. 
For example, it is doubtful web users now spend more time on amazonaws.com than on amazon.com.  According to a recent discussion from a CommonCrawl contributor~\citep{CCSampling}, the domain rank of CommonCrawl has around 33\% overlap with the general web traffic estimated by DNS providers.
In comparison, ClueWeb22 is designed to reflect the distribution of search engine traffic and its top websites align with our intuitions of search user behaviors.

\textbf{Differences in Available Information.} In ClueWeb22 we provide rich data affiliated with each web page, such as clean text, semantic annotations, and visual information. 
Some of them can be produced for CommonCrawl snapshots with a decent amount of effort and at varying quality. The most difficult one to obtain is the visual information. It requires using a web browser to render web pages and record their visual appearances, which is very engineering- and resource-heavy. 
This is perhaps one reason why such information, though being so valuable in industry systems, is overlooked in the research community. 

We highlight these differences not to show the superiority of either dataset, but to raise awareness of the underlying data properties for researchers and practitioners when using them.
All datasets are useful. Perhaps a more beneficial strategy is to utilize information from multiple resources. For example, one can leverage ClueWeb22 as the high quality source of web information and combine multiple months of CommonCrawl snapshots to increase the coverage on target scenarios, e.g., on some low-resource languages. The rich information from ClueWeb22 can also serve as weak supervision signals to train better content extraction systems for CommonCrawl.

\section{Related Corpora}
\label{sec:related}

ClueWeb22 is the third dataset in a lineage that began with ClueWeb09~\citep{ClueWeb09} and ClueWeb12~\citep{ClueWeb12}.  The previous datasets were constructed by Carnegie Mellon University using  open-source software (Apache Nutch~\citep{Nutch} in 2009 and Heritrix~\citep{Heritrix} in 2012). The crawling of web pages started from seed URLs and traversed the web graph by following hyperlinks.  ClueWeb09 contains 1 billion pages, 50\% English and 50\% in the next nine most popular languages on the Internet, roughly in proportion to their popularity.  
ClueWeb12 contains 730 million pages, primarily in English.  Both datasets were among the largest web corpora available when they were released.  They have been popular with researchers for more than a decade, but they are aging and no longer accurately represent the web of 2020s.

Many have created cleaner and more targeted web corpora by filtering multiple CommonCrawl snapshots. CCNet~\citep{wenzek2020ccnet} starts from the UTF-8 text of February 2019 CommonCrawl snapshot, performs de-duplication at text passage level,  identifies English texts, and keeps those closer to Wikipedia texts using a n-gram text classifier. It has been widely used to pretrain English language models, especially in the standard base++ and large++ RoBERTa-style models~\citep{liu2019roberta, meng2021coco}.
C4~\citep{raffel2019t5}, the pretraining corpus of T5, uses a series of language patterns to filter high  quality English texts from CommonCrawl~\citep{raffel2019t5}.
Other notable derivations from CommonCrawl include Pile-CC~\citep{gao2020pile}, which is constructed to pretrain GPT-Neo, and CC-News~\citep{Hamborg2017}, which targets news articles.

Another way to estimate the quality of web pages is by upvotes in online forum. For example, OpenAI scraped WebText by harvesting URLs with three or more upvotes on Reddit~\citep{radford2019language}. Similar approaches have been used to construct the publicly available OpenWebText~\citep{Gokaslan2019OpenWeb} and OpenWebText2~\citep{gao2020pile} corpora. 
These human-voted web pages likely consist of higher quality content than random web pages, but there is only a limited number of URLs covered by online forums.

Recently, places with access to large scale proprietary web corpora have constructed their own high quality web datasets to pretrain language models.
A series of recent large language models from Google used web pages sampled by a learned quality score from their web corpora~\citep{du2021glam, chowdhery2022palm}. DeepMind collected the MassiveText corpus to pretrain their large scale language models~\citep{rae2021scaling, hoffmann2022training}.
Their research explorations demonstrate the benefits and perhaps also necessity of large scale high quality web corpora in pretraining large neural models.
One goal of ClueWeb22 is to make this type of data resource available to the general research community.

\section{Licensing and Distribution}
\label{sec:licensing}

ClueWeb22 is licensed and distributed by Carnegie Mellon University and The Lemur Project using methods similar to those used for more than a decade for older ClueWeb datasets.  There is no fee to license the dataset, which is convenient for researchers who can access the dataset on a cloud computing platform that already has it.  There is a modest fee to cover the cost of distributing the dataset when it is necessary to transfer terabytes from Carnegie Mellon to a research organization, for example, the cost of hard disk and shipment.  
The Lemur Project website\footnote{\url{https://lemurproject.org/clueweb22/}} is kept current with information about updates, dataset maintenance, licensing, and distribution.

\section{Conclusions}
\label{sec:conclusion}

ClueWeb22 provides ten billion web pages sampled from the web discovered by crawlers of a commercial search engine. 
The sampling was conducted according to the preference of search engine users and includes web pages from the super head, head, and tail part of the search traffic. 
The included web pages are relatively clean with industry-quality filters to remove spam and adult content.
The result distribution of web pages in ClueWeb22 is as close as possible to the web distribution of commercial search in the real world.

Besides raw data, ClueWeb22 also includes rich information, such as browser-rendered web pages, visual information embedded in DOM trees, and semantic annotations from production-quality models.
These resources are widely used in various industry systems and critical for many search engine functions, but were hard to obtain without a commercial search product. 
Furthermore, to ease the barrier of entry, we pre-compute various data using these resources, including clean texts, document fields, and anchor graphs. These data serve as the standardized content of ClueWeb22. They can also be used as the starting point for more sophisticated and customized document understanding systems.

Access to large scale high quality web corpora becomes more and more important in AI research with the recent advancement of deep learning. 
We hope the release of ClueWeb22 reduces the boundary between research in places with proprietary data access and the general research community, thus empower more research progress in the near future.

\section{Acknowledgments}

We would like to thank the legal teams at Microsoft and Carnegie Mellon University for helping set up the licensing process, Ruohong Zhang for running the pretraining study, Junaid Ahmed for initializing the  idea of the ClueWeb22 effort together, and Likun Ouyang for dynamic rendering support to get high quality screenshots and visual features.

\bibliographystyle{ACM-Reference-Format}
\bibliography{sigirforum}
\end{document}